\documentclass[twocolumn,showpacs,eqsecnum,preprintnumbers,amsmath,amssymb,floatfix,prb]{revtex4}
\usepackage{graphicx}
\usepackage{dcolumn}
\usepackage{bm}

\begin{document}

\title{Field-theoretical renormalization group for a flat two-dimensional Fermi
surface}
\author{H. Freire, E. Corr\^{e}a and A. Ferraz}
 \affiliation{Laborat\'{o}rio de Supercondutividade,\\
 Centro Internacional de F\'{i}sica da Mat\'{e}ria Condensada,\\
 Universidade de Bras\'{i}lia - Bras\'{i}lia, Brazil}

\date{\today}

\begin{abstract}
We implement an explicit two-loop calculation of the coupling
functions and the self-energy of interacting fermions with a
two-dimensional flat Fermi surface in the framework of the field
theoretical renormalization group (RG) approach. Throughout the
calculation both the Fermi surface and the Fermi velocity are
assumed to be fixed and unaffected by interactions. We show that in
two dimensions, in a weak coupling regime, there is no significant
change in the RG flow compared to the well-known one-loop results
available in the literature. However, if we extrapolate the flow to
a moderate coupling regime there are interesting new features
associated with an anisotropic suppression of the quasiparticle
weight $Z$ along the Fermi surface, and the vanishing of the
renormalized coupling functions for several choices of the external
momenta.
\end{abstract}


\maketitle

\section{Introduction}

A better understanding of the physical properties of highly
interacting electrons in two spatial dimensions (2D) is believed to
be central for high-Tc superconductivity. Soon after the discovery
of the high-Tc superconductors, Anderson\cite{Anderson} suggested
that a strongly interacting 2D electron gas should resemble a 1D
Luttinger liquid state. This question remains unresolved to this
date. Thanks to the high precision angular resolved photoemission
experiments performed in a variety of materials\cite{Campuzano}, we
know, at present, important facts concerning the Fermi surface (FS)
of the cuprates. The FS of underdoped and optimally doped Bi2212 and
YBaCuO compounds contains both flat and curved sectors\cite{Dessau}.
As a result, they are nearly perfectly nested along certain
directions in momentum space. As is well-known, whenever there is a
flat FS, the corresponding one-electron dispersion law resembles a
1D dispersion.

The cuprates are Mott insulators at half-filling which become
metallic at very low doping\cite{Yoshida}. At half-filling,
Hubbard-like models have a square shape FS imposed by electron-hole
symmetry. The FS changes as we vary the filling factor and, as soon
as it is lightly doped, it acquires non-zero curvature sectors in
k-space. However, in the immediate vicinity of half-filling, there
are, at most, isolated curved spots in momentum space. Consequently,
in a first analysis, one may neglect their presence altogether.
Following this scheme, several workers investigated the properties
of a 2D electron gas in the presence of a totally flat
FS\cite{Ruvalds,Dzyaloshinskii,Doucot,Luther,Sudbo}. In their
approaches, the FS is always kept fixed and never deviates from its
original flat form. Besides that, their results conflict with each
other. Conventional perturbation theory calculations\cite{Ruvalds},
parquet method results\cite{Dzyaloshinskii}, as well as one-loop
perturbative RG calculations\cite{Doucot} indicate that, for
repulsive interactions, there is never a Luttinger liquid state in
2D. For the repulsive Hubbard interaction, the one-loop calculations
indicate that the antiferromagnetic spin density wave is the
dominant instability in two dimensions. In contrast, applying
bosonization methods, Luther was able to map the square FS onto two
sets of perpendicular chains\cite{Luther}. As a result of that, the
corresponding electron correlation functions become sums of power
law terms with exponents only differing in form from those of a
Luttinger liquid\cite{Sudbo}. Very recently, Rivasseau and
collaborators using a mathematically rigorous renormalization group
analysis\cite{Rivasseau} proved that, at finite temperatures, the
half-filled Hubbard model in two dimensions with a perfectly square
Fermi surface is indeed a Luttinger liquid apart from logarithmic
corrections. That result adds new elements to the possible existence
of a Luttinger liquid phase in two dimensions.

We report in this work a two-loop field theoretical RG calculation
for the electron gas in the presence of the same flat FS model as
used by Zheleznyak \emph{et al.}\cite{Dzyaloshinskii}.
Experimentally, such a FS was observed quite recently in an ARPES
experiment performed in $La_{2-x}Sr_{x}CuO_{4-\delta}$ (LSCO) thin
epitaxial film under strain\cite{Abrecht}. On the theoretical side,
all one-loop RG calculations presented so far for a flat 2D Fermi
surface indicate a flow to a strong coupling regime. One should
therefore check if a two-loop calculation changes that result or
not. Besides, there have been recent works addressing the
renormalization of the quasiparticle weight $Z$ due to
interactions\cite{Honerkamp,Katanin}. Those works are not totally
consistent since they ignore the feedback produced by $Z$ in the RG
equation for the coupling functions. We show that, as long as we
restrict ourselves to the weak coupling regime, $Z$ is only altered
slightly from its unity value. As a result, in that limit the RG
coupling flows are not severely modified by $Z$. In contrast, if we
consider moderate couplings, there are substantial changes produced
in the RG flow as opposed to the one-loop case, and $Z$ indeed goes
to zero in this region of coupling space.

In our analysis, we assume that the FS and the Fermi velocity
$v_{F}$ are not renormalized by the interactions. In a later work,
we show how the renormalization of both $v_{F}$ and the FS may
affect our results. The novel aspect of our work, apart from using a
field theoretical RG approach to deal with a 2D fermionic problem,
is the fact that we explicitly calculate all the two-loop diagrams
for the self-energy and the renormalized coupling functions, taking
full account of the variation of the momenta of the counterterms
along the FS. As a result, we are able to calculate the nonleading
logarithmic contributions originated by the so-called nonparquet
diagrams and the self-energy corrections which lead to a momentum
dependent quasiparticle weight $Z(\mathbf{p})$. To our knowledge, it
is the first time that such a systematic and detailed two-loop
calculation in two dimensions is presented in the context of the
fermionic RG approach. For simplicity, and to make easier the
comparison with other RG results, we use the so-called ``g-ology''
notation throughout this work. To reduce the number of accounted
diagrams, we limit ourselves to backscattering and forward
scattering processes only. The Umklapp scattering processes will not
be considered since they only occur when the distance between the
parallel FS patches is equal to $\pi$. We intend to include these
contributions in a later publication.

Since the fermionic field theoretical RG is not widely known among
condensed matter physicists we present the method at length and in
full detail. We begin by presenting our Lagrangian model and the
reasons why we need to apply the renormalization method in the first
place for the flat FS. We calculate next the one-particle
irreducible vertex functions associated with both forward and
backscattering processes in one-loop order. We derive the RG
equations for the renormalized coupling functions and show how they
flow to strong coupling in agreement with the parquet and the other
one-loop RG approaches. We then move on to calculate the self-energy
in two-loop order. Using the renormalization theory we determine the
quasiparticle weight $Z$ along the Fermi surface. Subsequently, we
add the nonleading contributions arising from the nonparquet
diagrams. As a result of that, we show numerically that the coupling
functions either flow to large plateau values or approach zero.
Those nonzero plateau values are however sensitive to our
discretization procedure. Therefore, we can neither establish their
stability nor completely rule out that the plateaus represent some
intermediate crossover regime associated with the inexact
discretization of the coupled integral equations. At the end, we
conclude by discussing how this picture should be altered by the
renormalization of the Fermi surface whose effects will be shown in
a subsequent work.

\section{Fermi Surface and Lagrangian Model}

\begin{figure}[t]
  \includegraphics[height=2.3in]{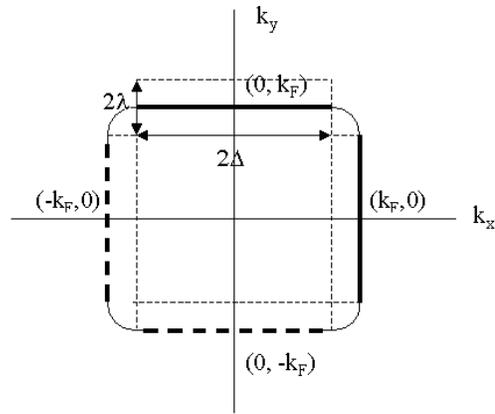}\\
  \caption{The 2D flat Fermi surface with rounded corners. We divide it into
  four regions: two of the solid line type and two of the dashed line type.
  The perpendicular regions do not mix in our scheme.}\label{sf}
\end{figure}

Our starting point is a strongly interacting 2D electron gas in the
presence of the flat FS shown schematically in Fig.1. In order to
keep a closer contact with well-known works in one-dimensional
physics\cite{Solyom}, we divide the FS into four regions. We
restrict the momenta at the FS to the flat parts only. The
contributions of the patches perpendicular to the $k_{x}$ and
$k_{y}$ directions do not mix with each other in our approach. For
convenience, we restrict ourselves to the one-electron states
labelled by the momenta $p_{\perp}=k_{y}$ and $p_{\parallel}=k_{x}$
associated with one of the two sets of perpendicular patches. The
momenta parallel to the FS are restricted to the interval
$-\Delta\leqslant p_{
\parallel}\leqslant\Delta$, where $\Delta$
essentially determines the size of the flat patches. The energy
dispersion of the single-particle states are given by
$\varepsilon_{a}\left(\mathbf{p}\right)=v_{F}\left(\left|p_{\perp}\right|-k_{F}\right)$
depending only on the momenta perpendicular to the Fermi surface.
The label $a=\pm$ refers to the flat sectors at $p_{\perp}=\pm
k_{F}$ respectively. Here we take
$k_{F}-\lambda\leqslant\left|p_{\perp}\right|\leqslant
k_{F}+\lambda$, where $\lambda$ is some fixed momentum cut-off. The
Fermi momentum $k_{F}$ is not renormalized in our approach and takes
its noninteracting value. We also neglect the Fermi velocity $v_{F}$
dependence along the FS. This is consistent with the fact that the
Fermi surface will not be renormalized in the present work.

The Lagrangian $L$ associated with the 2D flat Fermi surface is
given by
\begin{align}
L=&\sum_{\mathbf{p},\sigma,a=\pm}\psi_{(a)\sigma}^{\dagger}\left(\mathbf{p}\right)
\left[i\partial_{t}-v_{F}\left(\left|p_{\perp}\right|-k_{F}\right)\right]\psi_{(a)\sigma}
\left(\mathbf{p}\right)\nonumber
\\&-\frac{1}{V}\sum_{\mathbf{p,q,k}}\sum_{\alpha,\beta,\delta,\gamma}\left[g_{2}\delta_{\alpha\delta}\delta_{\beta\gamma}-g_{1}\delta_{\alpha\gamma}\delta_{\beta\delta}\right]\nonumber
\\&\times\psi_{\left(+\right)\delta}^{\dagger}\left(\mathbf{p+q-k}\right)\psi_{\left(-\right)\gamma}^{\dagger}\left(\mathbf{k}\right)
\psi_{\left(-\right)\beta}\left(\mathbf{q}\right)\psi_{\left(+\right)\alpha}\left(\mathbf{p}\right),\nonumber\\\label{lag1}
\end{align}

\begin{figure}[t]
  \includegraphics[width=3.0in]{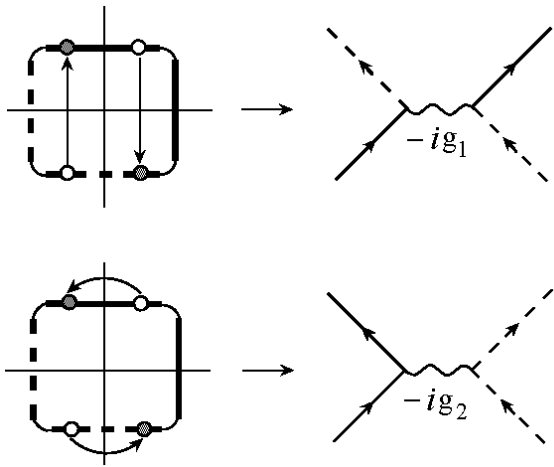}\\
  \caption{The interacion processes in the model and the corresponding Feynman rules for the vertices.
  The $g_{1}$ and $g_{2}$ couplings stand for backscattering and forward
  scattering respectively.}\label{processes}
\end{figure}

\noindent where we are following the ``g-ology'' notation. Here the
$\psi_{\left(\pm\right)}^{\dagger}$ and $\psi_{\left(\pm\right)}$
are, respectively, the creation and annihilation operators of
particles located at the $\pm$ patches. The couplings $g_{1}$ and
$g_{2}$ stand for backscattering and forward scattering. $V$ is the
volume of the two-dimensional system which, for convenience, will be
set equal to unity from this point on. The graphical representations
of the corresponding forward and backscattering interactions are
shown schematically in Fig.2. In all Feynman diagrams, the
single-particle propagators $G_{\left(+\right)}^{(0)}$ and
$G_{\left(-\right)}^{(0)}$ will be represented by a solid and a
dashed line respectively, according to their association with the
corresponding FS patches.

In setting up a conventional perturbation theory to do calculations
with this model one immediately encounters, in one-loop order,
logarithmic divergent particle-particle and particle-hole
diagrams\cite{Ferraz1,Ferraz2}
\begin{eqnarray}
\Pi^{(0)}\left(q_{0},q_{\perp}=0,q_{\parallel}\right)=
\int_{p}G_{\left(+\right)}^{\left(0\right)}\left(p\right)G_{\left(-\right)
}^{\left(0\right)}\left(-p+q\right),\label{pi}\nonumber \\ \\
\chi^{(0)}\left(q_{0},q_{\perp}=-2k_{F},q_{\parallel}\right)=
\int_{p}G_{\left(+\right)}^{\left(0\right)}\left(p\right)G_{\left(-\right)
}^{\left(0\right)}\left(p+q\right),\label{qui}\nonumber\\
\end{eqnarray}
where $p=(p_{0},p_{\perp},p_{\parallel})$ and
$\int_{p}=\int\frac{dp_{\parallel}}{2\pi}\frac{dp_{\perp}}{2\pi}\frac{dp_{0}}{2\pi}$.

Using the single-particle propagators from the non-interacting part
of the Lagrangian we find respectively
\begin{equation}
\Pi^{(0)}\left(q_{0}=\omega,q_{\parallel}\right)=i\frac{\left(2\Delta-\left|q_{\parallel}\right|\right)}{4\pi^{2}v_{F}}
\ln\left(\frac{\Omega}{\omega}\right),\label{pi1}
\end{equation}
and
\begin{equation}
\chi^{(0)}\left(q_{0}=\omega,q_{\parallel}\right)=-i\frac{\left(2\Delta-\left|q_{\parallel}\right|\right)}{4\pi^{2}v_{F}}
\ln\left(\frac{\Omega}{\omega}\right),\label{qui1}
\end{equation}
where $-\Delta\leqslant q_{\parallel}\leqslant\Delta$,
$\Omega=2v_{F}\lambda$ is a fixed upper energy cutoff and $\omega$
is an energy scale which goes to zero as we approach the Fermi
surface. These infrared divergent functions will appear infinitely
many times if we attempt to do a perturbative calculation of the
interacting one-particle Green's function
$G_{\left(\pm\right)}\left(p\right)$ or the effective two-particle
interaction, i.e. the one-particle irreducible function
$\Gamma^{\left(4\right)}\left(p_{1},p_{2},p_{3},p_{4}\right)$, to
infinite order. This will, of course, invalidate any conventional
perturbation theory approach to the problem and, at the same time,
will make meaningless the direct comparison of our results with
experiment. While the experiments find in principle finite values
for the measured physical quantities, our perturbative series
expansions are plagued with an infinite number of powers of infrared
logarithmic singularities.

We circumvent this problem following the field theory procedure of
introducing appropriate counterterms in the Lagrangian to render the
physical parameters finite in all scattering channels\cite{Peskin}.
In doing so, we eliminate the divergences order by order in
perturbation theory. What will become clear soon is that the 2D
Fermi surface will add important new features in this scheme. The
counterterms are now functions of momenta and vary continuously
along the FS. In the next section, we begin to show how to implement
the fermionic field theory RG in practice by calculating the
renormalized one-particle irreducible functions associated with both
backscattering and forward scattering up to one-loop order.

\section{Renormalized Coupling Functions at One-Loop Order}

\begin{figure}[t]
  \includegraphics[height=1.6in]{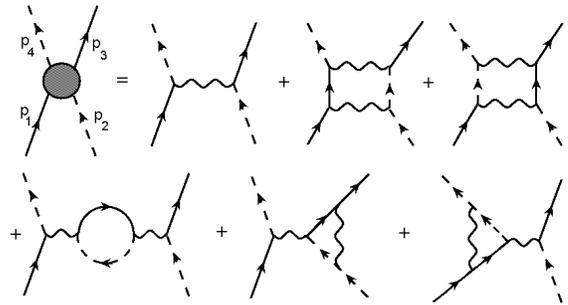}\\
  \caption{The Feynman diagrams up to one-loop order for the four-point vertex in the backscattering channel. The single
  particle propagators are represented by either solid or dashed lines according to their association with the corresponding
  FS patches.}\label{d11l}
\end{figure}

Let us calculate initially the one-particle irreducible function
$\Gamma^{\left(4\right)}$ associated with the backscattering channel
up to one-loop order. Using appropriate Feynman rules, we arrive at
\begin{eqnarray}
&&\Gamma_{1}^{\left(4\right)}\left(p_{1},p_{2},p_{3}\right)=-ig_{1}+2g_{1}g_{2}
\Pi^{(0)}(p_{1}+p_{2})\nonumber\\&&-2g_{1}^{2}\chi^{(0)}(p_{2}-p_{3})+2g_{1}g_{2}\chi^{(0)}(p_{2}-p_{3}).
\label{gama1}
\end{eqnarray}
where it will be implicitly assumed from this point on that
$p_{4}=p_{1}+p_{2}-p_{3}$ by momentum conservation. In addition, we
follow the same conventions as before concerning integrals and
energy-momenta representation. The diagrams corresponding to these
contributions are shown schematically in Fig.\ref{d11l}.

\begin{figure}[t]
  \includegraphics[height=0.75in]{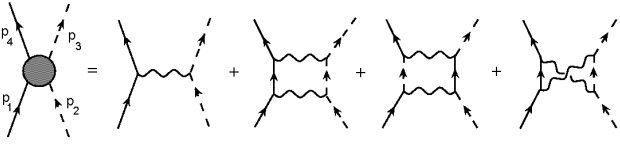}\\
  \caption{The Feynman diagrams up to one-loop order for the four-point vertex in the forward scattering channel.}\label{d21l}
\end{figure}

Let us next perform a similar calculation with the one-particle
irreducible function $\Gamma^{\left(4\right)}$ associated with the
forward scattering channel depicted in Fig.\ref{d21l}. Using once
again the Feynman rules associated with our initial $L$, we obtain
\begin{eqnarray}
&&\Gamma_{2}^{\left(4\right)}\left(p_{1},p_{2},p_{3}\right)=-ig_{2}+g_{1}^{2}
\Pi^{(0)}(p_{1}+p_{2})\nonumber\\&&+g_{2}^{2}\Pi^{(0)}(p_{1}+p_{2})+g_{2}^{2}\chi^{(0)}(p_{3}-p_{1}).\label{gama2}
\end{eqnarray}

\noindent It is clear from those two perturbative series expansions
that the internal particle-particle and particle-hole lines will
produce logarithmic singularities if the external momenta and
energies are located at the Fermi surface. To deal with this problem
and to regularize our perturbation expansions, we now invoke the
field theory method.

Being the effective two-particle interaction in the backscattering
channel, we define the finite one-particle irreducible
$\Gamma_{1}^{\left(4\right)}$ such that the renormalized
backscattering coupling function
$g_{1R}\left(p_{1\parallel},p_{2\parallel},p_{3\parallel};\omega\right)$
is given by
\begin{equation}
\Gamma_{1}^{\left(4\right)}\left(p_{1},p_{2},p_{3}\right)=-ig_{1R}\left(p_{1},p_{2},p_{3};\omega\right),
\label{gamapresc}
\end{equation}
where $p_{1\perp}=k_{F}$, $p_{2\perp}=-k_{F}$, $p_{3\perp}=k_{F}$,
$p_{4\perp}=-k_{F}$ and the corresponding energy components
$p_{10}=p_{20}=\omega/2$, $p_{30}=3\omega/2$ and $p_{40}=-\omega/2$.
Similarly for $p_{1\perp}=k_{F}$, $p_{2\perp}=-k_{F}$,
$p_{3\perp}=-k_{F}$, $p_{4\perp}=k_{F}$ and for
$p_{10}=p_{20}=\omega/2$, $p_{30}=3\omega/2$ and $p_{40}=-\omega/2$,
we relate the finite one-particle irreducible
$\Gamma_{2}^{\left(4\right)}$ and the renormalized forward coupling
function
$g_{2R}\left(p_{1\parallel},p_{2\parallel},p_{3\parallel};\omega\right)$
to each other

\begin{equation}
\Gamma_{2}^{\left(4\right)}\left(p_{1},p_{2},p_{3}\right)=-ig_{2R}\left(p_{1},p_{2},p_{3};\omega\right).
\label{gamapresc1}
\end{equation}

\noindent We now go back to our initial Lagrangian model and replace
the coupling constants $g_{1}$ and $g_{2}$ by the renormalized
coupling functions $g_{1R}(\{p_{i\parallel}\} ;\omega)$ and
$g_{2R}(\{p_{i\parallel}\} ;\omega)$ respectively. If we apply the
Feynman rules with our new forward and backscattering coupling
functions to calculate both $\Gamma_{1}^{\left(4\right)}$ and
$\Gamma_{2}^{\left(4\right)}$ at the Fermi surface we find instead

\begin{widetext}
\begin{align}
&\Gamma_{1}^{\left(4\right)} =
-ig_{1R}\left(p_{1\parallel},p_{2\parallel},p_{3\parallel}\right) +
\frac{i}{4\pi^{2}v_{F}}\bigg\{
\int_{\mathcal{D}_{1}}dk_{\parallel}\big[g_{1R}\left(-k_{\parallel}+p_{1\parallel}+
p_{2\parallel},k_{\parallel},p_{3\parallel}\right)g_{2R}\left(p_{1\parallel},p_{2\parallel},k_{\parallel}\right)\nonumber\\
&+g_{2R}\left(k_{\parallel},-k_{\parallel}+p_{1\parallel}+p_{2\parallel},p_{1\parallel}+p_{2\parallel}
-p_{3\parallel}\right)g_{1R}\left(p_{1\parallel},p_{2\parallel},k_{\parallel}\right)\big]
+\int_{\mathcal{D}_{2}}dk_{\parallel}\big[2g_{1R}\left(p_{1\parallel},p_{2\parallel}-p_{3\parallel}+k_{\parallel},k_{\parallel}\right)\nonumber\\
&\times
g_{1R}\left(k_{\parallel},p_{2\parallel},p_{3\parallel}\right)
-g_{1R}\left(p_{1\parallel},p_{2\parallel}-p_{3\parallel}+k_{\parallel},k_{\parallel}\right)
g_{2R}\left(k_{\parallel},p_{2\parallel},p_{2\parallel}-p_{3\parallel}+k_{\parallel}\right)
-g_{1R}\left(k_{\parallel},p_{2\parallel},p_{3\parallel}\right)\nonumber\\
&\times
g_{2R}\left(p_{1\parallel},p_{2\parallel}-p_{3\parallel}+k_{\parallel},p_{1\parallel}+p_{2\parallel}-p_{3\parallel}\right)
\big]\bigg\}\ln\left(\frac{\Omega}{\omega}\right),\label{gama11}
\end{align}
and
\begin{align}
&\Gamma_{2}^{\left(4\right)} =
-ig_{2R}\left(p_{1\parallel},p_{2\parallel},p_{3\parallel}\right) +
\frac{i}{4\pi^{2}v_{F}}\bigg\{
\int_{\mathcal{D}_{1}}dk_{\parallel}\big[g_{2R}\left(-k_{\parallel}+p_{1\parallel}+p_{2\parallel},k_{\parallel},p_{3\parallel}\right)g_{2R}\left(p_{1\parallel},p_{2\parallel},k_{\parallel}\right)\nonumber\\
&+g_{1R}\left(k_{\parallel},-k_{\parallel}+p_{1\parallel}+p_{2\parallel},p_{1\parallel}+p_{2\parallel}-p_{3\parallel}\right)g_{1R}\left(p_{1\parallel},p_{2\parallel},k_{\parallel}\right)\big]
-\int_{\mathcal{D}_{3}}dk_{\parallel}\big[g_{2R}\left(p_{1\parallel},p_{3\parallel}-p_{1\parallel}+k_{\parallel},p_{3\parallel}\right)\nonumber\\
&\times
g_{2R}\left(k_{\parallel},p_{2\parallel},p_{3\parallel}-p_{1\parallel}+k_{\parallel}\right)\big]\bigg\}\ln\left(\frac{\Omega}{\omega}\right),\label{gama21}
\end{align}
where all the domains of integration $\mathcal{D}_{i}$ are given
explicitly in Appendix A. As it stands, the
$\Gamma^{\left(4\right)}_{i}$'s are divergent as
$\omega\rightarrow0$, contradicting our earlier assumptions given in
Eqs.(\ref{gamapresc}) and (\ref{gamapresc1}). To remedy the
situation, we add a new term to $L$
\begin{eqnarray}
-\sum_{\mathbf{p,q,k}}\sum_{\alpha,\beta,\delta,\gamma}\left[\Delta
g_{2}\delta_{\alpha\delta}\delta_{\beta\gamma}-\Delta
g_{1}\delta_{\alpha\gamma}\delta_{\beta\delta}\right]
\psi_{\left(+\right)\delta}^{\dagger}\left(\mathbf{p+q-k}\right)\psi_{\left(-\right)\gamma}^{\dagger}\left(\mathbf{k}\right)
\psi_{\left(-\right)\beta}\left(\mathbf{q}\right)\psi_{\left(+\right)\alpha}\left(\mathbf{p}\right),
\label{count}
\end{eqnarray}
where
\begin{align}
&\Delta
g_{1R}\left(p_{1\parallel},p_{2\parallel},p_{3\parallel};\omega\right)=\frac{1}{4\pi^{2}v_{F}}\bigg\{
\int_{\mathcal{D}_{1}}dk_{\parallel}\big[g_{1R}\left(-k_{\parallel}+p_{1\parallel}+p_{2\parallel},k_{\parallel},p_{3\parallel}\right)
g_{2R}\left(p_{1\parallel},p_{2\parallel},k_{\parallel}\right)+g_{1R}\left(p_{1\parallel},p_{2\parallel},k_{\parallel}\right)\nonumber\\
&\times
g_{2R}\left(k_{\parallel},-k_{\parallel}+p_{1\parallel}+p_{2\parallel},p_{1\parallel}+p_{2\parallel}-p_{3\parallel}\right)
\big]+\int_{\mathcal{D}_{2}}dk_{\parallel}\big[2g_{1R}\left(p_{1\parallel},p_{2\parallel}-p_{3\parallel}+k_{\parallel},k_{\parallel}\right)
g_{1R}\left(k_{\parallel},p_{2\parallel},p_{3\parallel}\right)\nonumber
\\&-g_{1R}\left(p_{1\parallel},p_{2\parallel}-p_{3\parallel}+k_{\parallel},k_{\parallel}\right)
g_{2R}\left(k_{\parallel},p_{2\parallel},p_{2\parallel}-p_{3\parallel}+k_{\parallel}\right)
-g_{2R}\left(p_{1\parallel},p_{2\parallel}-p_{3\parallel}+k_{\parallel},p_{1\parallel}+p_{2\parallel}-p_{3\parallel}\right)
\nonumber\\
&\times
g_{1R}\left(k_{\parallel},p_{2\parallel},p_{3\parallel}\right)
\big]\bigg\}\ln\left(\frac{\Omega}{\omega}\right),\label{count1}
\end{align}
and
\begin{align}
&\Delta
g_{2R}\left(p_{1\parallel},p_{2\parallel},p_{3\parallel};\omega\right)=\frac{1}{4\pi^{2}v_{F}}\bigg\{
\int_{\mathcal{D}_{1}}dk_{\parallel}\big[g_{2R}\left(-k_{\parallel}+p_{1\parallel}+p_{2\parallel},k_{\parallel},p_{3\parallel}\right)
g_{2R}\left(p_{1\parallel},p_{2\parallel},k_{\parallel}\right)+g_{1R}\left(p_{1\parallel},p_{2\parallel},k_{\parallel}\right)\nonumber\\
&\times
g_{1R}\left(k_{\parallel},-k_{\parallel}+p_{1\parallel}+p_{2\parallel},p_{1\parallel}+p_{2\parallel}-p_{3\parallel}\right)
\big]
-\int_{\mathcal{D}_{3}}dk_{\parallel}\big[g_{2R}\left(p_{1\parallel},p_{3\parallel}-p_{1\parallel}+k_{\parallel},p_{3\parallel}\right)
\nonumber\\
&\times
g_{2R}\left(k_{\parallel},p_{2\parallel},p_{3\parallel}-p_{1\parallel}+k_{\parallel}\right)\big]\bigg\}\ln\left(\frac{\Omega}{\omega}\right).
\label{count2}
\end{align}
\end{widetext}

\begin{figure}[t]
  \includegraphics[width=3.0in]{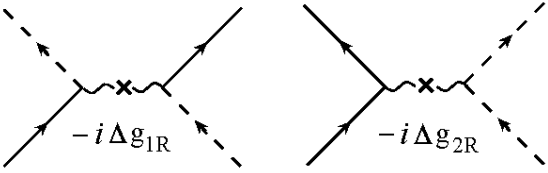}
  \caption{The additional Feynman rules for the counterterm vertices.}\label{counter0}
\end{figure}

The $\Delta g_{i}$'s are continuous functions of the scale $\omega$
and of the components of external momenta $p_{1\parallel}$,
$p_{2\parallel}$, $p_{3\parallel}$ only, since their related
$p_{i\perp}$ components are fixed at the Fermi surface. The presence
of these new terms in the Lagrangian results in additional Feynman
rules represented in Fig.\ref{counter0}.

We can adjust the counterterms such that all divergences are exactly
cancelled in our series expansion for $\Gamma_{1}^{\left(4\right)}$
and $\Gamma_{2}^{\left(4\right)}$ in one-loop order. But the price
we pay for this is the appearance of an energy scale $\omega$. All
physical quantities now depend on this new scale parameter. However,
the original theory does not know anything about $\omega$, i.e., the
initial parameters do not depend on the scale. This naturally leads
us to the renormalization group conditions for the renormalized
coupling functions
\begin{equation}
\omega\frac{d}{d\omega}\left(g_{1R}+\Delta
g_{1R}\right)=0,\label{rg11l}
\end{equation}
and

\begin{equation}
\omega\frac{d}{d\omega}\left(g_{2R}+\Delta
g_{2R}\right)=0.\label{rg21l}
\end{equation}

\noindent The RG flow equations for the renormalized coupling
functions follow immediately from this. We therefore find in
one-loop order
\begin{widetext}
\begin{align}
\omega\frac{dg_{1R}(p_{1\parallel},p_{2\parallel},p_{3\parallel})}{d\omega}&=\frac{1}{4\pi^{2}v_{F}}\bigg\{
\int_{\mathcal{D}_{1}}dk_{\parallel}\big[g_{1R}\left(-k_{\parallel}+p_{1\parallel}+p_{2\parallel},k_{\parallel},p_{3\parallel}\right)
g_{2R}\left(p_{1\parallel},p_{2\parallel},k_{\parallel}\right)\nonumber\\
&+g_{2R}\left(k_{\parallel},-k_{\parallel}+p_{1\parallel}+p_{2\parallel},p_{1\parallel}+p_{2\parallel}-p_{3\parallel}\right)
g_{1R}\left(p_{1\parallel},p_{2\parallel},k_{\parallel}\right)\big]\nonumber\\
&+\int_{\mathcal{D}_{2}}dk_{\parallel}\big[2g_{1R}\left(p_{1\parallel},p_{2\parallel}-p_{3\parallel}+k_{\parallel},k_{\parallel}\right)
g_{1R}\left(k_{\parallel},p_{2\parallel},p_{3\parallel}\right)\nonumber
\\&-g_{1R}\left(p_{1\parallel},p_{2\parallel}-p_{3\parallel}+k_{\parallel},k_{\parallel}\right)
g_{2R}\left(k_{\parallel},p_{2\parallel},p_{2\parallel}-p_{3\parallel}+k_{\parallel}\right)\nonumber\\
&-g_{2R}\left(p_{1\parallel},p_{2\parallel}-p_{3\parallel}+k_{\parallel},p_{1\parallel}+p_{2\parallel}-p_{3\parallel}\right)
g_{1R}\left(k_{\parallel},p_{2\parallel},p_{3\parallel}\right)\big]\bigg\},\label{g1r1l}\nonumber\\
\end{align}
and
\begin{align}
\omega\frac{dg_{2R}(p_{1\parallel},p_{2\parallel},p_{3\parallel})}{d\omega}&=\frac{1}{4\pi^{2}v_{F}}\bigg\{
\int_{\mathcal{D}_{1}}dk_{\parallel}\big[g_{2R}\left(-k_{\parallel}+p_{1\parallel}+p_{2\parallel},k_{\parallel},p_{3\parallel}\right)
g_{2R}\left(p_{1\parallel},p_{2\parallel},k_{\parallel}\right)\nonumber\\
&+g_{1R}\left(k_{\parallel},-k_{\parallel}+p_{1\parallel}+p_{2\parallel},p_{1\parallel}+p_{2\parallel}-p_{3\parallel}\right)
g_{1R}\left(p_{1\parallel},p_{2\parallel},k_{\parallel}\right)\big]\nonumber\\
&-\int_{\mathcal{D}_{3}}dk_{\parallel}\big[g_{2R}\left(p_{1\parallel},p_{3\parallel}-p_{1\parallel}+k_{\parallel},p_{3\parallel}\right)
g_{2R}\left(k_{\parallel},p_{2\parallel},p_{3\parallel}-p_{1\parallel}+k_{\parallel}\right)\big]\bigg\}.\label{g2r1l}\nonumber\\
\end{align}
\end{widetext}

\noindent These RG equations are exactly equal to the ones obtained
by Zheleznyak \emph{et al.}\cite{Dzyaloshinskii} using the parquet
method. In fact, performing one-loop RG calculations is equivalent
to sum the parquet class of diagrams up to infinite order.

Having shown in detail how the RG field theory operates in one-loop
order we can now proceed to calculate with this method the electron
self-energy up to two loops.

\section{Self-Energy and Quasiparticle Weight Up to Two Loops}

\begin{figure}[tr]
  \includegraphics[width=3.4in]{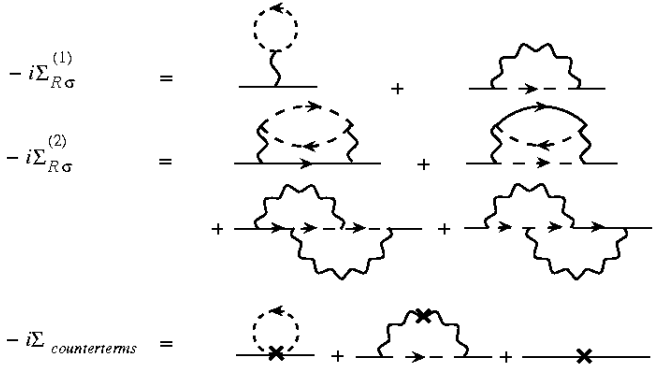}\\
  \caption{The diagrams for the self-energy calculation up to two-loop order.
  The last diagram of $-i\Sigma_{counterterms}$ is not computed at this stage. It represents
  the $\Delta Z$ contribution which will be added later on.}\label{selfened}
\end{figure}

The field theory RG eliminates the infrared divergences order by
order in the coupling functions. We do this by invoking the usual
Feynman rules at each order to obtain finite result in terms of the
renormalized physical quantities. Let us continue with the use of
this method to obtain the electron self-energy up to second order.
Using our already modified Feynman rules we have

\begin{eqnarray}
\Sigma_{R\sigma}\left(p\right)=\Sigma_{
R\sigma}^{\left(1\right)}+\Sigma_{R\sigma}^{\left(2\right)}
+\Sigma_{counterterms}. \label{selfen}
\end{eqnarray}

\noindent We show the diagrams corresponding to each of those
contributions in Fig.\ref{selfened}. The one-loop contributions
$-i\Sigma_{R\sigma}^{(1)}$ and the first two diagrams of
$-i\Sigma_{counterterms}$ produce a shift in $k_{F}$ and they are
crucial for the renormalization of the Fermi velocity and of the
Fermi surface itself\cite{Kopietz}. Those aspects will be discussed
in a subsequent paper. We neglect their contributions in the present
work altogether. Let us therefore concentrate our attention in the
two-loop terms $-i\Sigma_{R\sigma}^{(2)}$. We find
\begin{widetext}
\begin{align}
-i&\Sigma_{R\sigma}^{\left(2\right)}\left(p_{0},p_{\perp},p_{\parallel};\omega\right)
=\frac{i}{64\pi^{4}v_{F}^{2}}\left(p_{0}-v_{F}\left(p_{\perp}-k_{F}\right)\right)
\int_{\mathcal{D}_{4}}dk_{\parallel}
dq_{\parallel}[2g_{1R}\left(-k_{\parallel}+p_{\parallel}+q_{\parallel},k_{\parallel},q_{\parallel}\right)\nonumber\\
&\times
g_{1R}\left(p_{\parallel},q_{\parallel},k_{\parallel}\right)+
2g_{2R}\left(p_{\parallel},q_{\parallel},-k_{\parallel}+p_{\parallel}+q_{\parallel}\right)
g_{2R}\left(k_{\parallel},-k_{\parallel}+p_{\parallel}+q_{\parallel},q_{\parallel}\right)
-g_{1R}\left(p_{\parallel},q_{\parallel},k_{\parallel}\right)\nonumber \\
&\times
g_{2R}\left(k_{\parallel},-k_{\parallel}+p_{\parallel}+q_{\parallel},q_{\parallel}\right)
-g_{2R}\left(p_{\parallel},q_{\parallel},-k_{\parallel}+p_{\parallel}+q_{\parallel}\right)
g_{1R}\left(k_{\parallel},-k_{\parallel}+p_{\parallel}+q_{\parallel},p_{\parallel}\right)]\nonumber
\\
&\times\left[\ln\left(\frac{\Omega-v_{F}\left(p_{\perp}-k_{F}\right)-p_{0}-i\delta}{v_{F}\left(p_{\perp}-k_{F}\right)
-p_{0}-i\delta}\right)+\ln\left(\frac{\Omega-v_{F}\left(p_{\perp}-k_{F}\right)+p_{0}-i\delta}{v_{F}\left(p_{\perp}-k_{F}\right)
+p_{0}-i\delta}\right)\right].\label{selfen1}
\end{align}
\end{widetext}
where $\mathcal{D}_{4}$ is given in Appendix A. If we take the
limits $\omega\rightarrow0$ with $p_{\perp}=k_{F}$ or
$v_{F}\left(p_{\perp}-k_{F}\right)=\omega\rightarrow0$ at $p_{0}=0$
our perturbation theory produces an undesirable nonanalyticity at
the FS. The way to eliminate this is again to add a new term to our
Lagrangian of the form
\begin{equation}
F\left(p_{\parallel};\omega\right)\sum_{\mathbf{p},\sigma,a=\pm}\psi_{(a)\sigma}^{\dagger}\left(\mathbf{p}\right)
\left[i\partial_{t}-v_{F}\left(p_{\perp}-k_{F}\right)\right]\psi_{(a)\sigma}\left(\mathbf{p}\right).\label{count3}
\end{equation}
With this, the new ``noninteracting'' Lagrangian now reads
\begin{align}
&\sum_{\mathbf{p},\sigma,a=\pm}\left(1+F\left(p_{\parallel};\omega\right)\right)\nonumber \\
&\times\psi_{(a)\sigma}^{\dagger}\left(\mathbf{p}\right)
\left[i\partial_{t}-v_{F}\left(p_{\perp}-k_{F}\right)\right]\psi_{(a)\sigma}\left(\mathbf{p}\right).\label{count4}
\end{align}
Let us define $F\left(p_{\parallel};\omega\right)=\Delta
Z\left(p_{\parallel};\omega\right)=Z\left(p_{\parallel};\omega\right)-1$.
Now, in passing, we notice that the main effect of this new
counterterm is to renormalize the fermion field
\begin{equation}
\psi_{(a)\sigma}\left(\mathbf{p}\right)\rightarrow
Z^{\frac{1}{2}}\left(p_{\parallel};\omega\right)\psi_{(a)\sigma}\left(\mathbf{p}\right).\label{renorm}\end{equation}
To determine the function $\Delta
Z\left(p_{\parallel};\omega\right)$ we consider the inverse of the
renormalized one-particle Green's function
$\Gamma_{R\sigma}^{\left(2\right)}\left(p\right)$. Let us define
$\Gamma_{R\sigma}^{\left(2\right)}$ such that at $p_{\perp}=k_{F}$
and $p_{0}=\omega$
\begin{equation}
\Gamma_{R\sigma}^{\left(2\right)}\left(p_{0}=\omega,p_{\perp}=k_{F},p_{\parallel}\right)=\omega.\label{presc1}\end{equation}
Taking into account that
\begin{align}
\Gamma_{R\sigma}^{\left(2\right)}\left(p\right)=&Z\left(p_{\parallel};\omega\right)
\left[p_{0}-v_{F}\left(p_{\perp}-k_{F}\right)\right]\nonumber \\
&-\Sigma_{R\sigma}^{\left(2\right)}\left(p_{0},p_{\perp},p_{\parallel};\omega\right),\label{presc2}
\end{align}
it follows from our previous result that
\begin{widetext}
\begin{align}
&Z\left(p_{\parallel};\omega\right)=1-\frac{1}{32\pi^{4}v_{F}^{2}}
\int_{\mathcal{D}_{4}}dk_{\parallel}
dq_{\parallel}[2g_{1R}\left(-k_{\parallel}+p_{\parallel}+q_{\parallel},k_{\parallel},q_{\parallel}\right)
g_{1R}\left(p_{\parallel},q_{\parallel},k_{\parallel}\right)
+2g_{2R}\left(p_{\parallel},q_{\parallel},-k_{\parallel}+p_{\parallel}+q_{\parallel}\right)\nonumber
\\&\times g_{2R}\left(k_{\parallel},-k_{\parallel}+p_{\parallel}+q_{\parallel},q_{\parallel}\right)
-g_{1R}\left(p_{\parallel},q_{\parallel},k_{\parallel}\right)
g_{2R}\left(k_{\parallel},-k_{\parallel}+p_{\parallel}+q_{\parallel},q_{\parallel}\right)
-g_{2R}\left(p_{\parallel},q_{\parallel},-k_{\parallel}+p_{\parallel}+q_{\parallel}\right)\nonumber
\\&\times g_{1R}\left(k_{\parallel},-k_{\parallel}+p_{\parallel}+q_{\parallel},p_{\parallel}\right)]
\ln\left(\frac{\Omega}{\omega}\right) .\label{eqz}
\end{align}
\end{widetext}

\noindent This reproduces the result obtained earlier by Kishine and
Yonemitsu\cite{Kishine} within a Wilsonian RG framework. In addition
to that, making use of Eq.(\ref{eqz}) we obtain the following RG
equation for the quasiparticle weight \bigskip
\begin{equation}
\omega\frac{\partial
Z\left(p_{\parallel};\omega\right)}{\partial\omega}=\gamma
Z\left(p_{\parallel};\omega\right),\label{zrg}
\end{equation}
where the anomalous dimension $\gamma$ is given by
\begin{widetext}
\begin{align}
&\gamma\left(p_{\parallel}\right)=\frac{1}{32\pi^{4}v_{F}^{2}}\int_{\mathcal{D}_{4}}dk_{\parallel}
dq_{\parallel}[2g_{1R}\left(-k_{\parallel}+p_{\parallel}+q_{\parallel},k_{\parallel},q_{\parallel}\right)
g_{1R}\left(p_{\parallel},q_{\parallel},k_{\parallel}\right)
+2g_{2R}\left(p_{\parallel},q_{\parallel},-k_{\parallel}+p_{\parallel}+q_{\parallel}\right)\nonumber
\\&\times g_{2R}\left(k_{\parallel},-k_{\parallel}+p_{\parallel}+q_{\parallel},q_{\parallel}\right)
-g_{1R}\left(p_{\parallel},q_{\parallel},k_{\parallel}\right)
g_{2R}\left(k_{\parallel},-k_{\parallel}+p_{\parallel}+q_{\parallel},q_{\parallel}\right)
-g_{2R}\left(p_{\parallel},q_{\parallel},-k_{\parallel}+p_{\parallel}+q_{\parallel}\right)\nonumber
\\&\times g_{1R}\left(k_{\parallel},-k_{\parallel}+p_{\parallel}+q_{\parallel},p_{\parallel}\right)].\label{gama}
\end{align}
\end{widetext}

\begin{figure}[t]
  \includegraphics[height=1.7in]{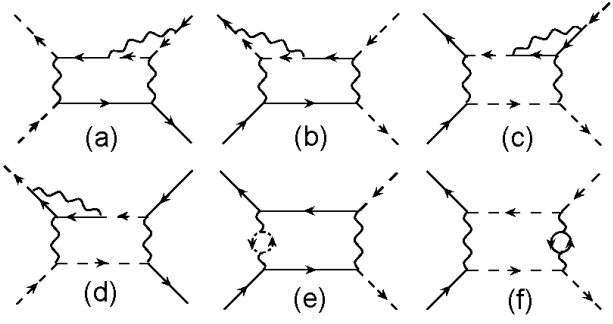}\\
  \caption{The nonparquet diagrams for the renormalized four-point vertex in the backscattering channel.}\label{nonparquet1}
\end{figure}

Before proceeding with the full calculation of the RG equations for
the renormalized coupling functions in two-loop order we call
attention to the fact that up to now, due to the addition of several
counterterms, our original Lagrangian can be written as

\bigskip

\begin{align}
&L=\sum_{\mathbf{p},\sigma,a=\pm}Z\psi_{(a)\sigma}^{\dagger}\left(\mathbf{p}\right)
\left[i\partial_{t}-v_{F}\left(p_{\perp}-k_{F}\right)\right]\psi_{(a)\sigma}\left(\mathbf{p}\right)\nonumber
\\&-\sum_{\mathbf{p,q,k}}\sum_{\alpha,\beta,\delta,\gamma}\left[\prod_{i=1}^{4}Z\left(p_{i\parallel}\right)\right]^{\frac{1}{2}}\left[
g_{2B}\delta_{\alpha\delta}\delta_{\beta\gamma}-
g_{1B}\delta_{\alpha\gamma}\delta_{\beta\delta}\right]\nonumber\\
&\times\psi_{\left(+\right)\delta}^{\dagger}\left(\mathbf{p+q-k}\right)\psi_{\left(-\right)\gamma}^{\dagger}\left(\mathbf{k}\right)
\psi_{\left(-\right)\beta}\left(\mathbf{q}\right)\psi_{\left(+\right)\alpha}\left(\mathbf{p}\right),
\label{lagr}
\end{align}

\bigskip

\noindent where $g_{1B}$ and $g_{2B}$ are the bare couplings which
are in turn defined to be
\begin{equation}
g_{iB}=\left[\prod_{i=1}^{4}Z\left(p_{i\parallel}\right)\right]^{-\frac{1}{2}}\left(g_{iR}+\Delta
g_{iR}\right).\label{coup}
\end{equation}
Together with the quasiparticle weight, they will render the theory
finite to all orders in perturbation theory as will become clear
soon.

\section{RG Equations at Two-Loop Order}

For convenience, let us rewrite our renormalized Lagrangian in the
form
\begin{align}
L&=\sum_{\mathbf{p},\sigma,a=\pm}\psi_{(a)\sigma}^{\dagger}\left(\mathbf{p}\right)
\left[i\partial_{t}-v_{F}\left(p_{\perp}-k_{F}\right)\right]\psi_{(a)\sigma}\left(\mathbf{p}\right)\nonumber
\\
&-\sum_{\mathbf{p,q,k}}\sum_{\alpha,\beta,\delta,\gamma}\left[
g_{2R}\delta_{\alpha\delta}\delta_{\beta\gamma}-
g_{1R}\delta_{\alpha\gamma}\delta_{\beta\delta}\right]\nonumber\\
&\times\psi_{\left(+\right)\delta}^{\dagger}\left(\mathbf{p+q-k}\right)\psi_{\left(-\right)\gamma}^{\dagger}\left(\mathbf{k}\right)
\psi_{\left(-\right)\beta}\left(\mathbf{q}\right)\psi_{\left(+\right)\alpha}\left(\mathbf{p}\right)\nonumber
\\
&+\sum_{\mathbf{p},\sigma,a=\pm}\Delta
Z\psi_{(a)\sigma}^{\dagger}\left(\mathbf{p}\right)
\left[i\partial_{t}-v_{F}\left(p_{\perp}-k_{F}\right)\right]\psi_{(a)\sigma}\left(\mathbf{p}\right)\nonumber
\\
&-\sum_{\mathbf{p,q,k}}\sum_{\alpha,\beta,\delta,\gamma}\left[\Delta
g_{2R}\delta_{\alpha\delta}\delta_{\beta\gamma}-\Delta
g_{1R}\delta_{\alpha\gamma}\delta_{\beta\delta}\right]\nonumber\\
&\times\psi_{\left(+\right)\delta}^{\dagger}\left(\mathbf{p+q-k}\right)\psi_{\left(-\right)\gamma}^{\dagger}\left(\mathbf{k}\right)
\psi_{\left(-\right)\beta}\left(\mathbf{q}\right)\psi_{\left(+\right)\alpha}\left(\mathbf{p}\right).\label{lag2l}
\end{align}

\begin{figure}[b]
  \includegraphics[width=3.3in]{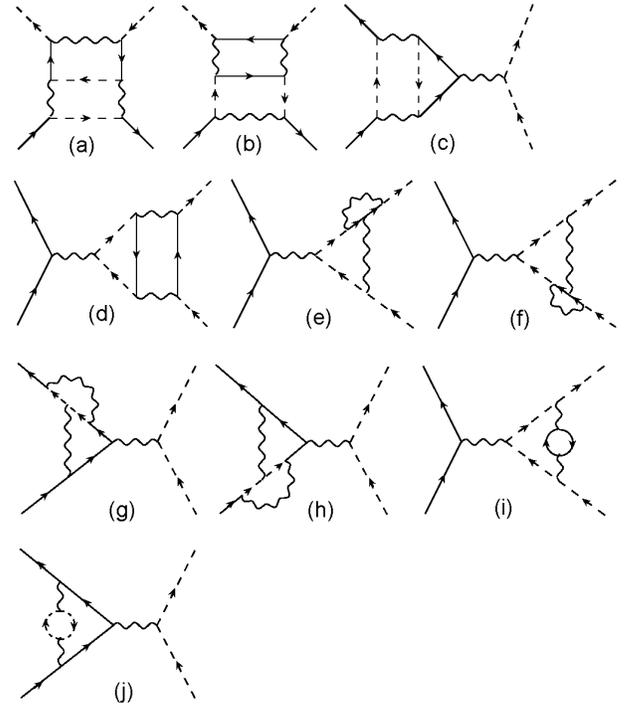}\\
  \caption{The nonparquet diagrams for the renormalized four-point vertex in the forward scattering channel.}\label{nonparquet2}
\end{figure}

\noindent Let us initially consider the diagrams obtained with our
new Feynman rules generated by this Lagrangian in two-loop order. At
this point, we only mention that to avoid double-counting of
diagrams we split up the counterterms into separate blocks. There
are several equivalence relations between diagrams generated by the
different constituent blocks. This will be explained further in
Appendix B. Taking all this into consideration, the contributions
due to the counterterms $\Delta g_{1R}$ and $\Delta g_{2R}$ cancel
exactly all the diagrams which are of the order of
$\ln^{2}\left(\Omega/\omega\right)$ in the vicinity of the Fermi
surface. As a result, in third-order, the only remaining
$\ln\left(\Omega/\omega\right)$ divergent contributions are produced
by the so-called nonparquet diagrams.

We display the nonparquet diagrams for $g_{1R}$ in Fig.7. In view of
that, to be consistent with the RG condition given in
Eq.(\ref{gamapresc}), we have therefore to redefine $\Delta g_{1R}$
using the same values for energies and momentum components
perpendicular to FS as before. We now find
\begin{align}
&\Delta
g_{1R}\left(p_{1\parallel},p_{2\parallel},p_{3\parallel}\right)=\Delta
g_{1R}^{1-loop} +\Delta g_{1R(a)}^{2-loop}+\Delta
g_{1R(b)}^{2-loop}\nonumber \\&+\Delta g_{1R(c)}^{2-loop}+\Delta
g_{1R(d)}^{2-loop}+\Delta g_{1R(e)}^{2-loop}+\Delta
g_{1R(f)}^{2-loop},\label{count5}
\end{align}
where the $\Delta g_{1R(x)}^{2-loop}$, with $x = a ... f$, are given
explicitly in Appendix A.

We can follow the same procedure to determine $\Delta g_{2R}$ in
two-loop order. The nonparquet diagrams for $g_{2R}$ are displayed
in Fig.8. As before, we now get
\begin{align}
&\Delta
g_{2R}\left(p_{1\parallel},p_{2\parallel},p_{3\parallel}\right)=\Delta
g_{2R}^{1-loop}+\Delta g_{2R(a)}^{2-loop}+\Delta
g_{2R(b)}^{2-loop}\nonumber \\&+\Delta g_{2R(c)}^{2-loop}+\Delta
g_{2R(d)}^{2-loop}+\Delta g_{2R(e)}^{2-loop}+\Delta
g_{2R(f)}^{2-loop}\nonumber
\\&+\Delta g_{2R(g)}^{2-loop}+\Delta
g_{2R(h)}^{2-loop}+\Delta g_{2R(i)}^{2-loop}+\Delta
g_{2R(j)}^{2-loop},\label{count6}
\end{align}
where the $\Delta g_{2R(x)}^{2-loop}$, with $x = a ... j$, are given
explicitly in Appendix A.

With the expressions of both $\Delta g_{1R}$ and $\Delta g_{2R}$ up
to two-loop order we can proceed with the calculation of the
accompanying RG equations for the $g_{iR}$'s. Taking into account
the associated $Z$ factors for the external momenta and the RG
conditions for the bare coupling functions $dg_{iB}/d\omega=0$
together with Eq.(\ref{gama}) we obtain immediately the two-loop RG
equations
\begin{eqnarray}
\omega\frac{dg_{iR}\left(p_{1\parallel},p_{2\parallel},p_{3\parallel}\right)}{d\omega}=&&\frac{1}{2}\sum_{j=1}^{4}\gamma\left(p_{j\parallel}\right)g_{iR}\left(p_{1\parallel},p_{2\parallel},p_{3\parallel}\right)
\nonumber \\&&-\omega\frac{\partial\Delta
g_{iR}\left(p_{1\parallel},p_{2\parallel},p_{3\parallel}\right)}{\partial\omega}.
\label{gir2l}
\end{eqnarray}

\section{Numerical Results}

Since it is impossible to solve all these RG equations analytically,
we have to resort to numerical methods in order to estimate how the
coupling functions change as we vary the scale $\omega$ to take the
physical system towards the FS. Here the basic idea is to discretize
the FS continuum replacing the interval $-\Delta \leqslant
p_{\parallel} \leqslant \Delta$ by a discrete set of 33 points. For
convenience, we use $\omega = \Omega\exp(-l)$, where $\Omega =
2k_{F}\lambda$ with $l$ being our RG step. We will choose
$\Omega/\Delta=1$. In view of our choice of points for the FS, we
are only allowed to go up to $l\approx2.8$ in the RG flow to avoid
the distance $\omega$ to the FS being smaller than the distance
between points since the discretization procedure no longer applies
in this case.

The choice of the initial conditions at $l=0$ in the RG equations
are, in principle, arbitrary. This is related to the fact that one
may choose any microscopic model to start with in order to study its
low-energy properties. Since we are most interested in relating our
results to a repulsive Hubbard model in 2D we initially set the
couplings equal to a Hubbard on-site repulsive interaction parameter
$U$ in our numerical scheme. As we will see shortly, the RG flows
are very sensitive to this initial value. The physical nature of the
resulting state is influenced directly by that choice.

In order to show this, at a first stage, we choose
$\overline{g}_{1R}=\overline{g}_{2R}=10$, where
$\overline{g}_{iR}=g_{iR}/\pi v_{F}$. Here, we apply the standard
fourth-order Runge-Kutta method to calculate the flow of the
couplings $\overline{g}_{1R}$ and $\overline{g}_{2R}$
self-consistently. We will present our results for three cases. The
first one is the one-loop approach. Next, we move on to a somewhat
intermediate step which includes the contributions of the
self-energy or the Z factor feedback into the one-loop RG equations
and analyze the corresponding flows. Then we present the full
two-loop order RG approach, which includes, in addition to the
self-energy feedback, the so-called nonparquet diagram contributions
in the RG equations. We will show that in this final case the RG
flows are very different from the one-loop case in view of the fact
that the quasiparticle weight $Z$ becomes very strongly suppressed
as we approach the FS.

Later on, we will change the initial condition, choosing a lower
value for the renormalized couplings in order to contrast it with
the previous choice. We take
$\overline{g}_{1R}=\overline{g}_{2R}=1$. We will see in this case
that since the quasiparticle weight does not change much from its
unity initial value, the feedback effect of that factor into the RG
equations is not very drastic. As a result, the two-loop flows will
resemble very much the corresponding one-loop flow for this
particular choice of initial interaction strength.

\begin{figure}[t]
  \includegraphics[width=3.3in]{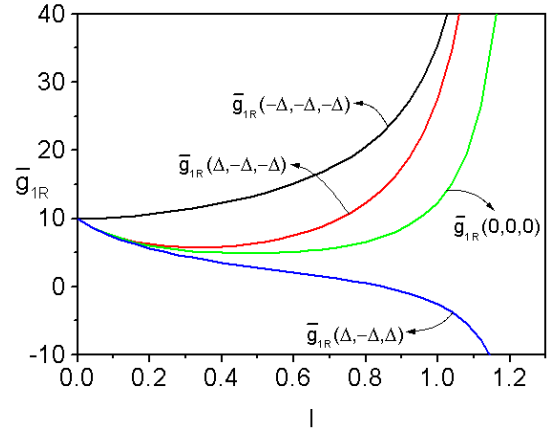}\\
  \caption{The one-loop RG flow of
  $\overline{g}_{1R}(p_{1\parallel},p_{2\parallel},p_{3\parallel})$
  for some choices of momenta .}\label{g1numr1l}
\end{figure}

\begin{figure}[b]
  \includegraphics[width=3.3in]{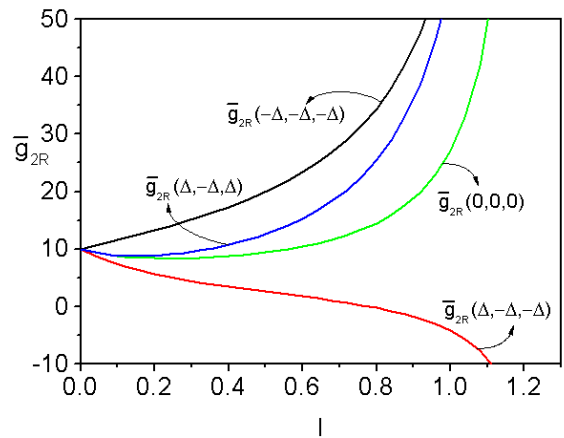}\\
  \caption{The one-loop RG flow of
  $\overline{g}_{2R}(p_{1\parallel},p_{2\parallel},p_{3\parallel})$
  for some choices of momenta .}\label{g2numr1l}
\end{figure}

\subsection{One-loop RG approach}

The one-loop results are depicted in Figs.\ref{g1numr1l} and
\ref{g2numr1l}. In them, we observe that even though the couplings
are initially constant (in our case, equal to ten), they acquire a
distinct momentum dependence as we approach the FS. In addition,
there are no fixed points in the flow, and all coupling values go
unequivocally to a strong coupling regime.

This result is consistent with other approaches based on a one-loop
order perturbative expansion such as the so-called parquet method
and other RG schemes, which predict an instability towards an
insulating spin density wave state with no sign of nonconventional
metallic behavior ever to be found in the physical
system\cite{Dzyaloshinskii}.

However, in all those approaches, the quasiparticle weight Z is
always equal to unity throughout the calculation. This should be
contrasted with the situation in which, due to interactions, the
quasiparticle weight might approach zero as well. To discuss what
happens in this case, we move on to the next step, which is the
inclusion of the Z factor feedback into the RG flow equations.
\begin{figure}[t]
  \includegraphics[width=3.3in]{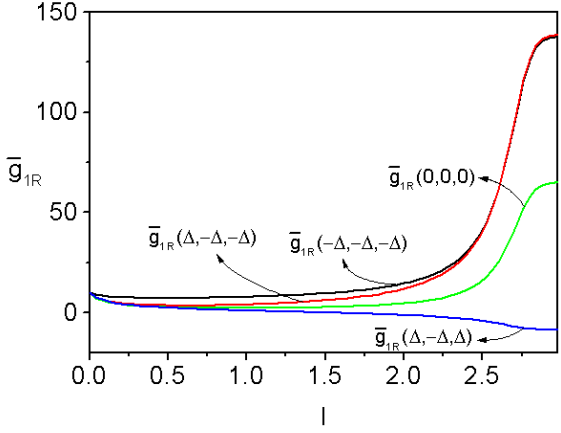}\\
  \caption{The one-loop with the Z factor RG flow of
  $\overline{g}_{1R}(p_{1\parallel},p_{2\parallel},p_{3\parallel})$
  for some choices of momenta.}\label{g1numrz}
\end{figure}

\begin{figure}[b]
  \includegraphics[width=3.3in]{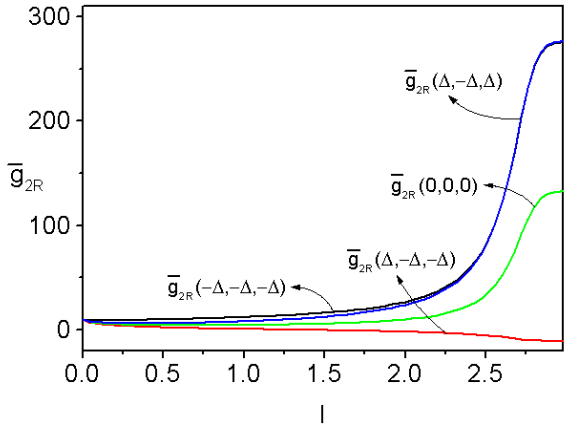}\\
  \caption{The one-loop with the Z factor RG flow of
  $\overline{g}_{2R}(p_{1\parallel},p_{2\parallel},p_{3\parallel})$
  for some choices of momenta.}\label{g2numrz}
\end{figure}

\subsection{One-loop with the Z factor approach}

For this case, we depict the results in Figs.\ref{g1numrz} and
\ref{g2numrz}. In that intermediate approach, which was also
discussed by Kishine and Yonemitsu\cite{Kishine}, we observe that
the initial tendency from one-loop analysis for the couplings to go
to a strong coupling regime is preserved. However, as the
suppression of $Z$ becomes prominent the flows seem to stop at fixed
values. In view of the fact that those values are sensitive to our
discretization procedure, we can not associate those results with
the existence of stable fixed points. If one increases the number of
points necessary for discretizing the FS by some factor, the fixed
values (or the plateaus in the figures) increase approximately by
the same factor. They should reach a higher bound regardless of the
number of points in the discretization procedure in order to be
characterized as true fixed points of our RG approach.

Since this last approach is not yet a complete two-loop calculation
anyway, we move on to the calculation of the full two-loop RG
equations for our problem.

\subsection{The full two-loop approach}

\begin{figure}[b]
  \includegraphics[width=3.3in]{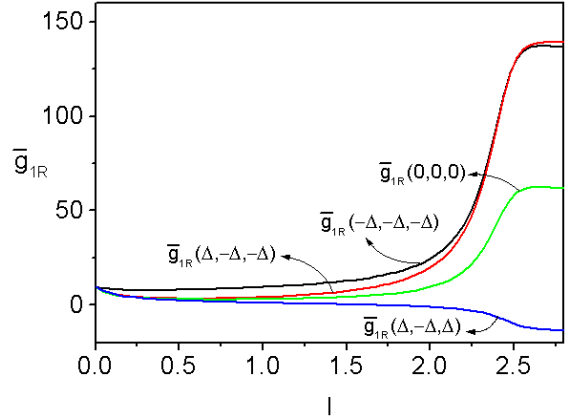}\\
  \caption{The full two-loop RG flow of
  $\overline{g}_{1R}(p_{1\parallel},p_{2\parallel},p_{3\parallel})$
  for some choices of momenta.}\label{g1numr2l}
\end{figure}

\begin{figure}[t]
  \includegraphics[width=3.3in]{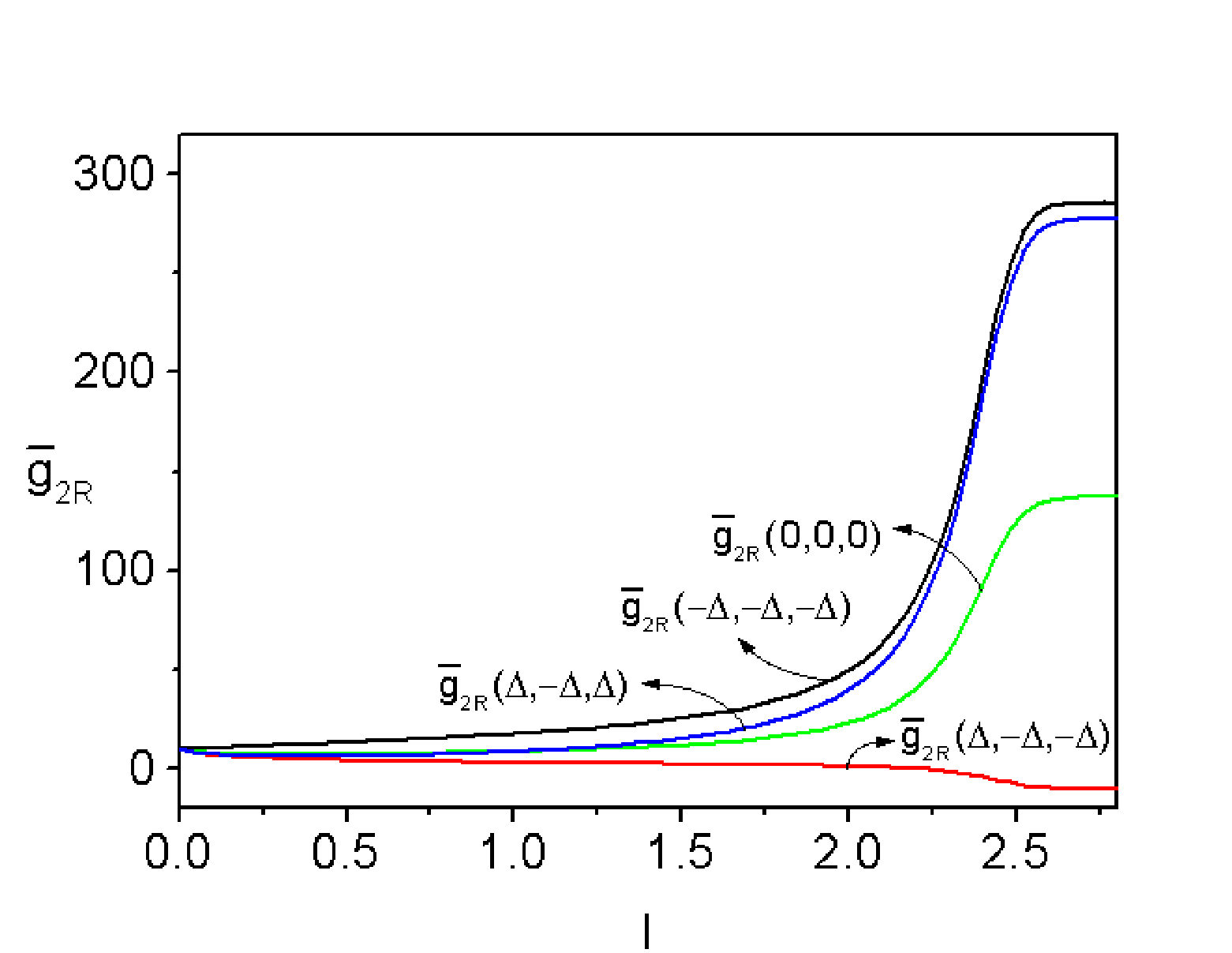}\\
  \caption{The full two-loop RG flow of
  $\overline{g}_{2R}(p_{1\parallel},p_{2\parallel},p_{3\parallel})$
  for some choices of momenta.}\label{g2numr2l}
\end{figure}

\begin{figure}[b]
  \includegraphics[height=2.7in]{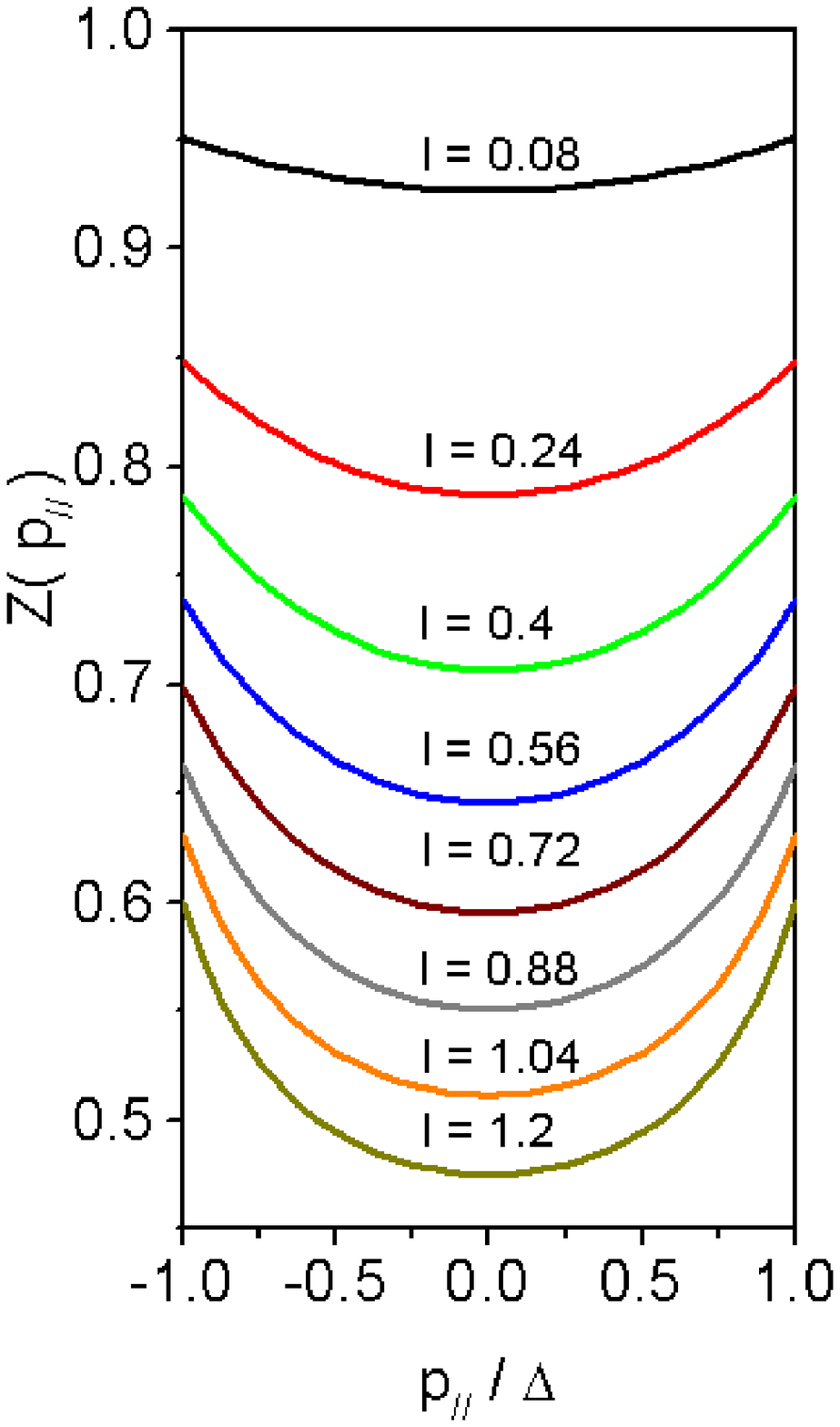}
  \includegraphics[height=2.7in]{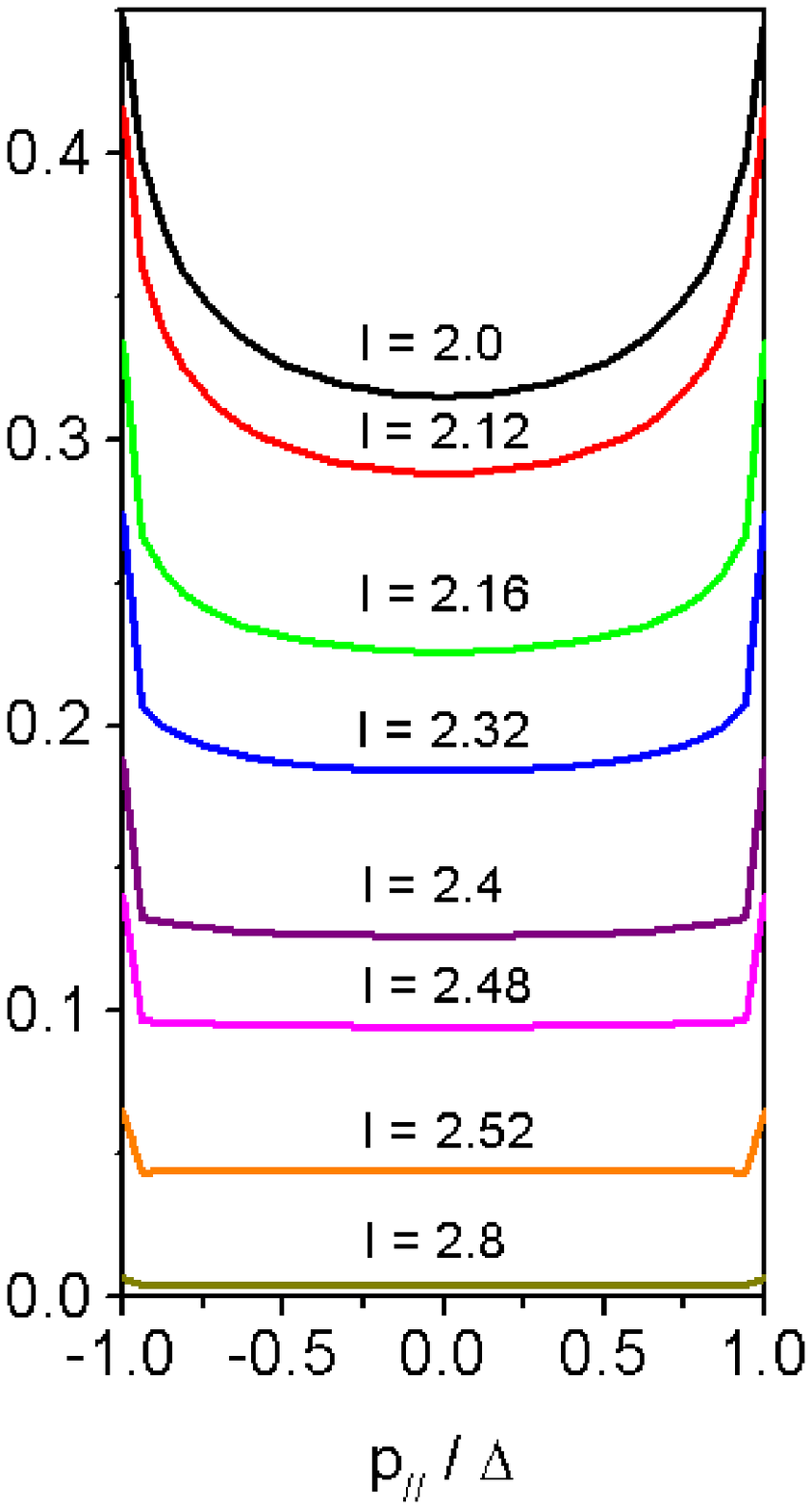}
  \caption{The RG flow for the quasiparticle weight as a function of the momentum $p_{\parallel}$
  along the FS for $\overline{g}_{1R}(l=0)=\overline{g}_{2R}(l=0)=10$.}\label{zppar}
\end{figure}

\begin{figure}[t]
  \includegraphics[width=3.1in]{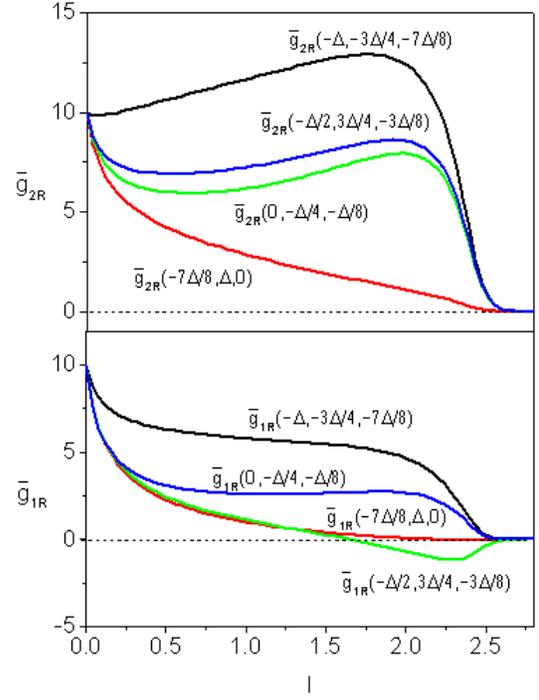}\\
  \caption{The vanishing of
  $\overline{g}_{2R}(p_{1\parallel},p_{2\parallel},p_{3\parallel})$ and
  $\overline{g}_{1R}(p_{1\parallel},p_{2\parallel},p_{3\parallel})$
  for some choices of momenta.}\label{g1g20}
\end{figure}

\begin{figure}[b]
  \includegraphics[height=2.6in]{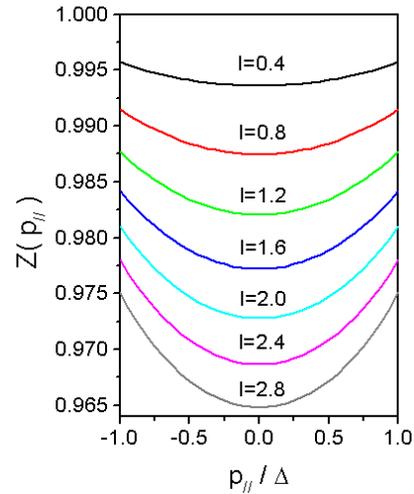}\\
  \caption{The RG flow for the quasiparticle weight as a function of the momentum $p_{\parallel}$
  along the FS for $\overline{g}_{1R}(l=0)=\overline{g}_{2R}(l=0)=1$.}\label{geq1}
\end{figure}

In this final case, we solve Eq.(\ref{gir2l}) numerically including
both the self-energy feedback and the nonparquet diagram
contributions for two different choices of initial conditions
$\overline{g}_{1R}=\overline{g}_{2R}=10$, and
$\overline{g}_{1R}=\overline{g}_{2R}=1$.

For the first choice, the results are shown in Figs.\ref{g1numr2l}
and \ref{g2numr2l}. In these figures, we observe that the RG flows,
at least qualitatively, do not change much from the previous case.
Again, we verify the initial one-loop tendency to flow to a strong
coupling regime, but as soon as the $Z$ effect becomes important the
rate of change of the couplings reduces abruptly. However, the
renormalized coupling functions do not saturate as in the previous
case (the above-mentioned plateaus). Although the couplings seem to
reach a plateau regime they now in fact change their values
continuously at a very slow rate.

Since these two-loop results are directly associated with the
effects produced by the quasiparticle weight we might as well
analyze the flow as we approach the FS. In order to do that, we must
also calculate Eq.(\ref{zrg}) self-consistently using the same
method described above. This equation is solved numerically assuming
that for $l=0$ the quasiparticle weight $Z(l=0)=1$. In doing that,
we observe numerically a rapid and very anisotropic suppression of
$Z$ for a moderate coupling regime when we approach the FS (see
Fig.\ref{zppar}). The quasiparticle weight vanishes faster at the
center parts of the square FS, i.e. in the $(\pi,0)$ or
$(0,\pi)$-directions. This is in qualitative agreement with Kishine
and Yonemitsu\cite{Kishine}, although they ignore the nonparquet
contributions and hence they do not perform a full two-loop
calculation.

Besides the rapid suppression of the quasiparticle weight $Z$, the
two-loop flow of the coupling functions displays other crucial
differences in contrast with the one-loop case. While in the
one-loop order all couplings seem to increase indefinitely, this is
certainly not the case in two-loop. There exists also many couplings
which go to zero continuously as we approach the FS. In fact,
whenever
$\overline{g}_{i}\left(p_{1\parallel},p_{2\parallel},p_{3\parallel}\right)$
is such that $p_{1\parallel}\neq p_{2\parallel}\neq p_{3\parallel}$,
the couplings renormalize to zero and are irrelevant in the RG
sense. In Fig.\ref{g1g20}, we show some couplings which belong to
this class.

Now, let us move on to the second choice of initial condition for
the couplings, i.e. $\overline{g}_{1R}=\overline{g}_{2R}=1$. Since
we already know that the distinct two-loop features are directly
related to what happens to the quasiparticle weight, we will limit
our RG analysis to the investigation of the $Z$ flow in this case.
As we see from Fig.\ref{geq1}, $Z$ is only mildly different from
unity in the RG flow. As a result, the corresponding two-loop flows
are very similar to the one-loop flow in this case. In order words,
as expected, for initial weak coupling strengths only, it is a
reasonable approximation to keep $Z=1$ fixed throughout the
calculation.

Therefore, we observe that models with different physical initial
coupling strengths may produce ultimately very different physical
states. In our numerical estimates, there is a clear indication of
crossover behavior between the regimes defined either by a strongly
suppressed quasiparticle weight or a $Z$ nearly equal to unity when
we vary the initial values of the couplings.

\section{Conclusion}

In summary, we showed in full detail a two-loop field theoretical RG
calculation for a square and fixed two-dimensional FS with rounded
corners. We neglected Umklapp effects at this stage since they only
occur when the distance between the parallel FS patches is equal to
$\pi$. We derived explicitly the Feynman rules of the renormalized
Lagrangian which regularize all the divergences generated by
perturbation theory. We did that by constructing appropriate
counterterms order by order in perturbation theory. Contrary to what
happens in local field theory models, those counterterms are now
functions of momenta varying continuously along the Fermi surface.
We showed that our method reproduces the results of other RG schemes
in one-loop order with all the coupling functions flowing
unequivocally to strong coupling. This behavior is typical of the
dominant spin-density wave instability in that regime, which signals
antiferromagnetic ordering as well as the Mott insulating condition.

Then, we moved on to the calculation of the self-energy correction
at two-loop order. Using appropriate assumptions about the form of
the inverse of the interacting one-particle Green's function at the
FS, we were able to calculate the quasiparticle weight Z, which was
shown to vary continuously along the FS. Next, we calculated the RG
equations for the coupling functions, taking into account both the
leading order term and the feedback of Z into the equations
themselves. Although the initial tendency observed in the one-loop
analysis of flowing to a strong coupling regime remained, the flows
then stopped suddenly at fixed plateau values. Despite that, we
emphasized that this was not a clear cut evidence of fixed points in
our approach since those values are sensitive to the number of
points used in the discretization procedure.

Finally, we proceeded with the calculation of the full two-loop
order RG equations for the renormalized coupling functions. In order
to do that, we had to calculate also the nonleading divergences
which arose from the so-called nonparquet diagrams. As we have seen,
their contribution was not cancelled out by the diagrams generated
by the counterterms $\Delta g_{1R}$ and $\Delta g_{2R}$ defined in
one-loop order. That automatically led us to redefine the
counterterms which from then on acquired also third-order
contributions. For this final case, we chose two different initial
conditions to analyze the flows, namely
$\overline{g}_{1R}=\overline{g}_{2R}=10$ and
$\overline{g}_{1R}=\overline{g}_{2R}=1$.

For the first choice of initial conditions, the flows were
qualitatively similar to the ones which included the feedback of
$Z$. They were, as a consequence, very dissimilar to the one-loop
flow. Afterwards, we numerically analyzed the corresponding flow of
the quasiparticle weight as we approached the FS. We observed that
$Z$ was anisotropic and strongly suppressed in that limit. We
pointed out that $Z$ vanishes faster in the $(\pi,0)$ or
$(0,\pi)$-directions, i.e the center parts of the square FS.
Finally, we emphasized that, in view of this strong suppression,
there were many coupling functions that renormalized to zero in
contrast to the one-loop case where they appeared to increase
indefinitely.

In contrast, for the second choice of initial conditions
($\overline{g}_{1R}=\overline{g}_{2R}=1$), $Z$ changes very mildly
from its original unity value. Consequently, the two-loop flows were
very similar to the one-loop ones. That is an indication, verified
numerically, that the physical state depends directly on the initial
choice of coupling strengths. In view of that, we intend to explore
further the implications of these results. In addition, we must also
analyze how the renormalization of the Fermi velocity and the FS
affect our results. This is even more important in view of the
recent results obtained by Afchain \emph{et al.}\cite{Rivasseau}
concerning the existence of a Luttinger liquid state in two spatial
dimensions. As we already noticed from our self-energy at one-loop
order, the Fermi surface and the Fermi velocity are directly
affected by interactions. The flows of the FS and $v_{F}$ to
critical regimes naturally impose important new conditions on the
flow of the renormalized coupling functions. The coupled RG
equations for the renormalized coupling functions, the FS and Fermi
velocity should all be solved self-consistently. That naturally
places extra difficulties but one should, nevertheless, deal with
this problem seriously. We plan to consider those questions in a
subsequent publication\cite{Hermann}.

\begin{acknowledgments}
We wish to acknowledge innumerous discussions with several
colleagues during the ``RG Methods for Interacting Electrons''
Winter School which took place in the ICCMP - UnB, Bras\'{i}lia,
during July $12^{th}$ - August $6^{th}$, 2004. We also wish to thank
Dionys Baeriswyl and Peter Kopietz for useful comments and
suggestions. This work was partially supported by the Financiadora
de Estudos e Projetos (FINEP) and by the Conselho Nacional de
Desenvolvimento Cient\'{i}fico e Tecnol\'{o}gico (CNPq) - Brazil.
\end{acknowledgments}

\appendix

\section{}

In this appendix, we will write down the explicit form of the
nonparquet contributions which are taken into account in the Eqs.
(\ref{count5}) and (\ref{count6}). We also give the several
intervals of integration along the Fermi surface that are considered
throughout this work which are the following
\begin{center}%
\begin{displaymath}
\mathcal{D}_{1}=\left\{%
\begin{array}{ll}
    -\Delta\leqslant k_{\parallel} \leqslant \Delta, & \hbox{} \\
    -\Delta\leqslant p_{1\parallel}+p_{2\parallel}-k_{\parallel} \leqslant \Delta. & \hbox{} \\
\end{array}%
\right.
\end{displaymath}
\begin{displaymath}
\mathcal{D}_{2}=\left\{%
\begin{array}{ll}
    -\Delta\leqslant k_{\parallel} \leqslant \Delta, & \hbox{} \\
    -\Delta\leqslant k_{\parallel}+p_{2\parallel}-p_{3\parallel} \leqslant \Delta. & \hbox{} \\
\end{array}%
\right.
\end{displaymath}
\begin{displaymath}
\mathcal{D}_{3}=\left\{%
\begin{array}{ll}
    -\Delta\leqslant k_{\parallel} \leqslant \Delta, & \hbox{} \\
    -\Delta\leqslant k_{\parallel}+p_{3\parallel}-p_{1\parallel} \leqslant \Delta. & \hbox{} \\
\end{array}%
\right.
\end{displaymath}
\begin{displaymath}
\mathcal{D}_{4}=\left\{%
\begin{array}{ll}
    -\Delta\leqslant k_{\parallel} \leqslant \Delta, & \hbox{} \\
    -\Delta\leqslant q_{\parallel} \leqslant \Delta, & \hbox{} \\
    -\Delta\leqslant p_{\parallel} \leqslant \Delta, & \hbox{} \\
    -\Delta\leqslant -k_{\parallel}+p_{\parallel}+q_{\parallel} \leqslant \Delta. & \hbox{} \\
\end{array}%
\right.
\end{displaymath}
\begin{displaymath}
\mathcal{D}_{5}=\left\{%
\begin{array}{ll}
    -\Delta\leqslant k_{\parallel} \leqslant \Delta, & \hbox{} \\
    -\Delta\leqslant q_{\parallel} \leqslant \Delta, & \hbox{} \\
    -\Delta\leqslant k_{\parallel}+q_{\parallel}-p_{1\parallel} \leqslant \Delta, & \hbox{} \\
    -\Delta\leqslant k_{\parallel}+p_{2\parallel}-p_{3\parallel} \leqslant \Delta. & \hbox{} \\
\end{array}%
\right.
\end{displaymath}
\begin{displaymath}
\mathcal{D}_{6}=\left\{%
\begin{array}{ll}
    -\Delta\leqslant k_{\parallel} \leqslant \Delta, & \hbox{} \\
    -\Delta\leqslant q_{\parallel} \leqslant \Delta, & \hbox{} \\
    -\Delta\leqslant k_{\parallel}+q_{\parallel}-p_{2\parallel} \leqslant \Delta, & \hbox{} \\
    -\Delta\leqslant k_{\parallel}+p_{3\parallel}-p_{2\parallel} \leqslant \Delta. & \hbox{} \\
\end{array}%
\right.
\end{displaymath}
\begin{displaymath}
\mathcal{D}_{7}=\left\{%
\begin{array}{ll}
    -\Delta\leqslant k_{\parallel} \leqslant \Delta, & \hbox{} \\
    -\Delta\leqslant q_{\parallel} \leqslant \Delta, & \hbox{} \\
    -\Delta\leqslant k_{\parallel}+q_{\parallel}-p_{4\parallel} \leqslant \Delta, & \hbox{} \\
    -\Delta\leqslant k_{\parallel}+p_{3\parallel}-p_{1\parallel} \leqslant \Delta. & \hbox{} \\
\end{array}%
\right.
\end{displaymath}
\begin{displaymath}
\mathcal{D}_{8}=\left\{%
\begin{array}{ll}
    -\Delta\leqslant k_{\parallel} \leqslant \Delta, & \hbox{} \\
    -\Delta\leqslant q_{\parallel} \leqslant \Delta, & \hbox{} \\
    -\Delta\leqslant k_{\parallel}+q_{\parallel}-p_{3\parallel} \leqslant \Delta, & \hbox{} \\
    -\Delta\leqslant k_{\parallel}+p_{1\parallel}-p_{3\parallel} \leqslant \Delta. & \hbox{} \\
\end{array}%
\right.
\end{displaymath}
\end{center}
In addition to that, we must have of course
\begin{center}
\begin{eqnarray}
    -\Delta\leqslant p_{1\parallel} \leqslant \Delta, \nonumber\\
    -\Delta\leqslant p_{2\parallel} \leqslant \Delta, \nonumber\\
    -\Delta\leqslant p_{3\parallel} \leqslant \Delta, \nonumber\\
    -\Delta\leqslant p_{4\parallel} \leqslant \Delta. \nonumber\\
    \nonumber
\end{eqnarray}%
\end{center}

We begin with the nonparquet diagrams shown in
Fig.(\ref{nonparquet1}) associated with $\Gamma_{1}^{(4)}$ we get
\begin{align}
&\Delta
g_{1R(a)}^{2-loop}=\frac{1}{32\pi^{4}v_{F}^{2}}\ln\left(\frac{\Omega}{\omega}\right)\nonumber
\\ &\times \int_{\mathcal{D}_{8}} d{k_{\parallel}dq_{\parallel}}g_{1R}
\left(k_{\parallel}+p_{1\parallel}-p_{3\parallel},p_{2\parallel},k_{\parallel}\right)\nonumber
\\&\times g_{1R}\left(p_{1\parallel},k_{\parallel}+q_{\parallel}-p_{3\parallel},k_{\parallel}+p_{1\parallel}-p_{3\parallel}\right)\nonumber
\\
&\times
g_{2R}\left(k_{\parallel},q_{\parallel},k_{\parallel}+q_{\parallel}-p_{3\parallel}
\right),\label{count7}
\end{align}

\begin{align}
\Delta
g_{1R(b)}^{2-loop}=\frac{1}{32\pi^{4}v_{F}^{2}}&\ln\left(\frac{\Omega}{\omega}\right)
\int_{\mathcal{D}_{8}}
d{k_{\parallel}dq_{\parallel}}g_{1R}\left(k_{\parallel},q_{\parallel},p_{3\parallel}\right)\nonumber
\\&\times g_{1R}
\left(k_{\parallel}+p_{1\parallel}-p_{3\parallel},p_{2\parallel},k_{\parallel}\right)\nonumber
\\ &\times g_{2R}\left(p_{1\parallel},k_{\parallel}+q_{\parallel}-p_{3\parallel},q_{\parallel}
\right),\label{count8}
\end{align}

\begin{align}
\Delta
g_{1R(c)}^{2-loop}=\frac{1}{32\pi^{4}v_{F}^{2}}&\ln\left(\frac{\Omega}{\omega}\right)
\int_{\mathcal{D}_{7}}
d{k_{\parallel}dq_{\parallel}}g_{2R}\left(q_{\parallel},k_{\parallel},p_{4\parallel}\right)\nonumber
\\&\times g_{1R}
\left(p_{1\parallel},k_{\parallel}+p_{3\parallel}-p_{1\parallel},p_{3\parallel}\right)\nonumber
\\ &\times g_{1R}\left(k_{\parallel}+q_{\parallel}-p_{4\parallel},p_{2\parallel},q_{\parallel}
\right),\label{count9}
\end{align}

\begin{align}
&\Delta
g_{1R(d)}^{2-loop}=\frac{1}{32\pi^{4}v_{F}^{2}}\ln\left(\frac{\Omega}{\omega}\right)\nonumber
\\ &\times \int_{\mathcal{D}_{7}} d{k_{\parallel}dq_{\parallel}}g_{1R}
\left(p_{1\parallel},k_{\parallel}+p_{3\parallel}-p_{1\parallel},p_{3\parallel}\right)\nonumber
\\&\times g_{2R}\left(k_{\parallel}+q_{\parallel}-p_{4\parallel},p_{2\parallel},k_{\parallel}+p_{3\parallel}-p_{1\parallel}
\right)\nonumber
\\
&\times
g_{1R}\left(q_{\parallel},k_{\parallel},k_{\parallel}+q_{\parallel}-p_{4\parallel}\right),\label{count10}
\end{align}

\begin{align}
&\Delta
g_{1R(e)}^{2-loop}=-\frac{1}{16\pi^{4}v_{F}^{2}}\ln\left(\frac{\Omega}{\omega}\right)\nonumber
\\ &\times \int_{\mathcal{D}_{8}} d{k_{\parallel}dq_{\parallel}}g_{1R}
\left(k_{\parallel}+p_{1\parallel}-p_{3\parallel},p_{2\parallel},k_{\parallel}\right)\nonumber
\\&\times g_{2R}\left(p_{1\parallel},k_{\parallel}+q_{\parallel}-p_{3\parallel},q_{\parallel}
\right)\nonumber
\\
&\times
g_{2R}\left(k_{\parallel},q_{\parallel},k_{\parallel}+q_{\parallel}-p_{3\parallel}\right),\label{count11}
\end{align}
and finally
\begin{align}
\Delta g_{1R(f)}^{2-loop}&=-\frac{1}{16\pi^{4}v_{F}^{2}}
\int_{\mathcal{D}_{7}}
d{k_{\parallel}dq_{\parallel}}g_{2R}\left(k_{\parallel},q_{\parallel},p_{4\parallel}\right)\nonumber
\\&\times g_{2R}
\left(k_{\parallel}+q_{\parallel}-p_{4\parallel},p_{2\parallel},k_{\parallel}+p_{3\parallel}-p_{1\parallel}\right)\nonumber
\\&\times g_{1R}\left(p_{1\parallel},k_{\parallel}+p_{3\parallel}-p_{1\parallel},p_{3\parallel}
\right)\ln\left(\frac{\Omega}{\omega}\right),\label{count12}
\end{align}
using the same conventions for the integrals over momenta components
along the Fermi surface as before.

Similarly for the nonparquet diagrams shown in
Fig.(\ref{nonparquet2}) associated with $\Gamma_{2}^{(4)}$ we get
\begin{align}
&\Delta
g_{2R(a)}^{2-loop}=\frac{1}{32\pi^{4}v_{F}^{2}}\ln\left(\frac{\Omega}{\omega}\right)\nonumber
\\ &\times \int_{\mathcal{D}_{5}} d{k_{\parallel}dq_{\parallel}}g_{1R}
\left(p_{1\parallel},k_{\parallel}+q_{\parallel}-p_{1\parallel},k_{\parallel}\right)\nonumber
\\&\times g_{1R}\left(k_{\parallel},p_{2\parallel},k_{\parallel}+p_{2\parallel}-p_{3\parallel}
\right)\nonumber
\\
&\times
g_{1R}\left(k_{\parallel}+p_{2\parallel}-p_{3\parallel},q_{\parallel},p_{4\parallel}\right),\label{count13}
\end{align}

\begin{align}
\Delta g_{2R(b)}^{2-loop}&=\frac{1}{32\pi^{4}v_{F}^{2}}
\int_{\mathcal{D}_{6}}
d{k_{\parallel}dq_{\parallel}}g_{1R}\left(p_{1\parallel},k_{\parallel},p_{4\parallel}\right)\nonumber
\\&\times g_{1R}\left(q_{\parallel},k_{\parallel}+p_{3\parallel}-p_{2\parallel},k_{\parallel}+q_{\parallel}-p_{2\parallel}
\right)\nonumber
\\&\times g_{1R}
\left(k_{\parallel}+q_{\parallel}-p_{2\parallel},p_{2\parallel},q_{\parallel}\right)\ln\left(\frac{\Omega}{\omega}\right),\label{count14}
\end{align}

\begin{align}
\Delta g_{2R(c)}^{2-loop}&=-\frac{1}{16\pi^{4}v_{F}^{2}}
\int_{\mathcal{D}_{5}}
d{k_{\parallel}dq_{\parallel}}g_{2R}\left(k_{\parallel},p_{2\parallel},p_{3\parallel}\right)\nonumber
\\&\times g_{1R}\left(p_{1\parallel},k_{\parallel}+q_{\parallel}-p_{1\parallel},k_{\parallel}
\right)\nonumber
\\&\times g_{1R}
\left(k_{\parallel}+p_{2\parallel}-p_{3\parallel},q_{\parallel},p_{4\parallel}\right)\ln\left(\frac{\Omega}{\omega}\right),\label{count15}
\end{align}

\begin{align}
&\Delta
g_{2R(d)}^{2-loop}=-\frac{1}{16\pi^{4}v_{F}^{2}}\ln\left(\frac{\Omega}{\omega}\right)\nonumber
\\ &\times \int_{\mathcal{D}_{6}} d{k_{\parallel}dq_{\parallel}}g_{2R}
\left(p_{1\parallel},k_{\parallel},k_{\parallel}+p_{3\parallel}-p_{2\parallel}\right)\nonumber
\\&\times g_{1R}\left(q_{\parallel},k_{\parallel}+p_{3\parallel}-p_{2\parallel},k_{\parallel}+q_{\parallel}-p_{2\parallel}\right)\nonumber
\\
&\times
g_{1R}\left(k_{\parallel}+q_{\parallel}-p_{2\parallel},p_{2\parallel},q_{\parallel}
\right),\label{count16}
\end{align}

\begin{align}
&\Delta
g_{2R(e)}^{2-loop}=\frac{1}{32\pi^{4}v_{F}^{2}}\ln\left(\frac{\Omega}{\omega}\right)\nonumber
\\ &\times \int_{\mathcal{D}_{6}} d{k_{\parallel}dq_{\parallel}}g_{2R}
\left(p_{1\parallel},k_{\parallel},k_{\parallel}+p_{3\parallel}-p_{2\parallel}\right)\nonumber
\\&\times g_{1R}\left(q_{\parallel},k_{\parallel}+p_{3\parallel}-p_{2\parallel},k_{\parallel}+q_{\parallel}-p_{2\parallel}\right)\nonumber
\\
&\times
g_{2R}\left(k_{\parallel}+q_{\parallel}-p_{2\parallel},p_{2\parallel},k_{\parallel}
\right),\label{count17}
\end{align}

\begin{align}
&\Delta
g_{2R(f)}^{2-loop}=\frac{1}{32\pi^{4}v_{F}^{2}}\ln\left(\frac{\Omega}{\omega}\right)\nonumber
\\ &\times \int_{\mathcal{D}_{6}} d{k_{\parallel}dq_{\parallel}}g_{2R}
\left(p_{1\parallel},k_{\parallel},k_{\parallel}+p_{3\parallel}-p_{2\parallel}\right)\nonumber
\\&\times g_{1R}\left(k_{\parallel}+q_{\parallel}-p_{2\parallel},p_{2\parallel},q_{\parallel}\right)\nonumber
\\
&\times
g_{2R}\left(q_{\parallel},k_{\parallel}+p_{3\parallel}-p_{2\parallel},p_{3\parallel}
\right),\label{count18}
\end{align}

\begin{align}
\Delta g_{2R(g)}^{2-loop}&=\frac{1}{32\pi^{4}v_{F}^{2}}
\int_{\mathcal{D}_{5}}
d{k_{\parallel}dq_{\parallel}}g_{2R}\left(k_{\parallel},p_{2\parallel},p_{3\parallel}\right)\nonumber
\\&\times g_{1R}\left(k_{\parallel}+p_{2\parallel}-p_{3\parallel},q_{\parallel},p_{4\parallel}
\right)\nonumber
\\&\times g_{2R}
\left(p_{1\parallel},k_{\parallel}+q_{\parallel}-p_{1\parallel},q_{\parallel}\right)\ln\left(\frac{\Omega}{\omega}\right),\label{count19}
\end{align}

\begin{align}
\Delta g_{2R(h)}^{2-loop}&=\frac{1}{32\pi^{4}v_{F}^{2}}
\int_{\mathcal{D}_{5}}
d{k_{\parallel}dq_{\parallel}}g_{2R}\left(k_{\parallel},p_{2\parallel},p_{3\parallel}\right)\nonumber
\\&\times g_{2R}\left(k_{\parallel}+p_{2\parallel}-p_{3\parallel},q_{\parallel},k_{\parallel}+q_{\parallel}-p_{1\parallel}
\right)\nonumber
\\&\times g_{1R}
\left(p_{1\parallel},k_{\parallel}+q_{\parallel}-p_{1\parallel},k_{\parallel}\right)\ln\left(\frac{\Omega}{\omega}\right),\label{count20}
\end{align}

\begin{align}
&\Delta
g_{2R(i)}^{2-loop}=-\frac{1}{16\pi^{4}v_{F}^{2}}\ln\left(\frac{\Omega}{\omega}\right)\nonumber
\\ &\times \int_{\mathcal{D}_{6}} d{k_{\parallel}dq_{\parallel}}g_{2R}
\left(p_{1\parallel},k_{\parallel},k_{\parallel}+p_{3\parallel}-p_{2\parallel}\right)\nonumber
\\&\times g_{2R}\left(k_{\parallel}+q_{\parallel}-p_{2\parallel},p_{2\parallel},k_{\parallel}\right)\nonumber
\\
&\times
g_{2R}\left(q_{\parallel},k_{\parallel}+p_{3\parallel}-p_{2\parallel},p_{3\parallel}
\right),\label{count21}
\end{align}
and finally
\begin{figure}[b]
  \includegraphics[width=2.8in]{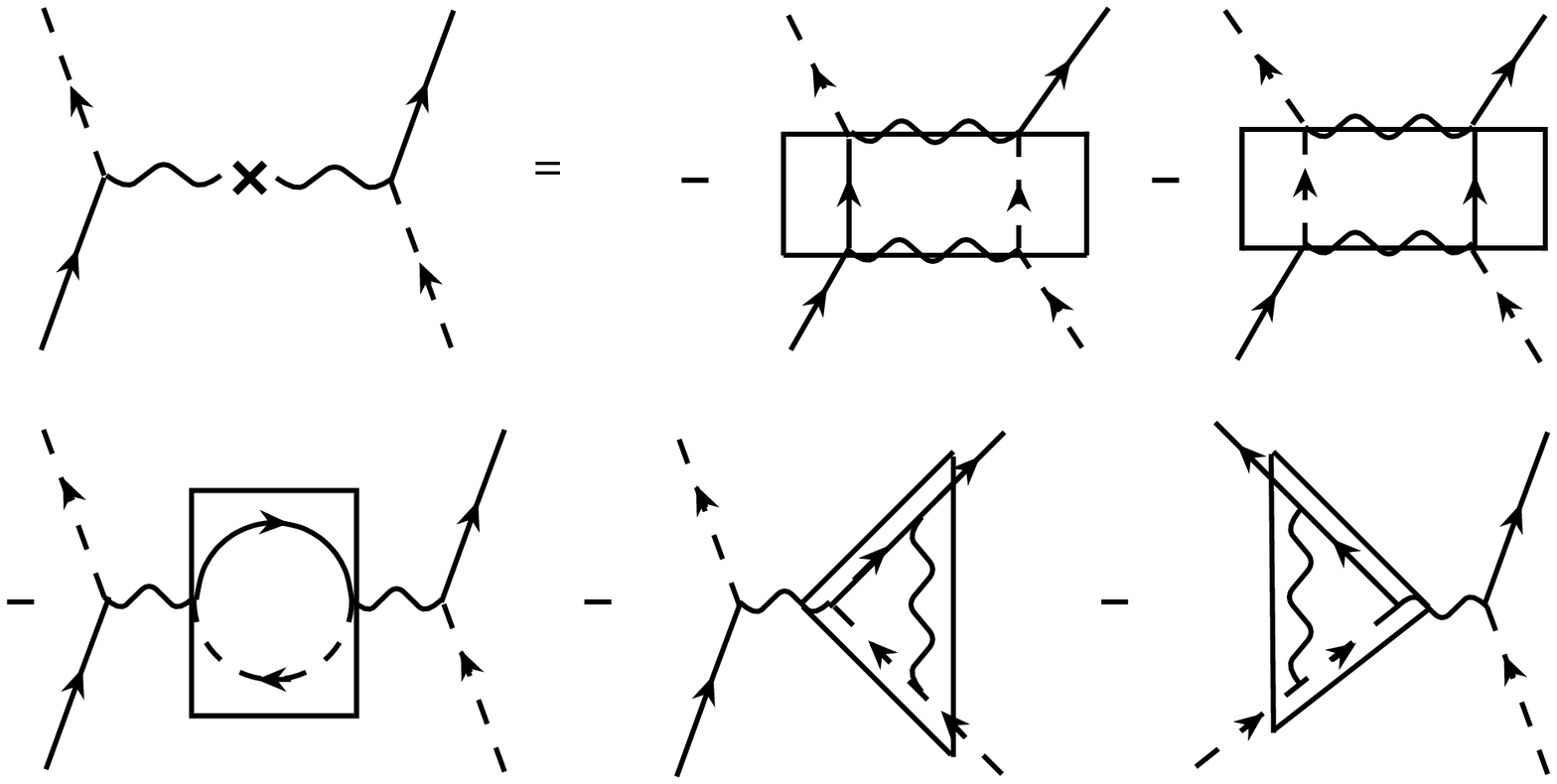}
  \caption{The counterterm diagrams of $\Delta g_{1R}$ up to one-loop order displayed in terms of constituent blocks.}\label{counter}
\end{figure}
\begin{align}
\Delta g_{2R(j)}^{2-loop}&=-\frac{1}{16\pi^{4}v_{F}^{2}}
\int_{\mathcal{D}_{5}}
d{k_{\parallel}dq_{\parallel}}g_{2R}\left(k_{\parallel},p_{2\parallel},p_{3\parallel}\right)\nonumber
\\&\times g_{2R}\left(k_{\parallel}+p_{2\parallel}-p_{3\parallel},q_{\parallel},k_{\parallel}+q_{\parallel}-p_{1\parallel}
\right)\nonumber
\\&\times g_{2R}
\left(p_{1\parallel},k_{\parallel}+q_{\parallel}-p_{1\parallel},q_{\parallel}\right)\ln\left(\frac{\Omega}{\omega}\right).\label{count22}
\end{align}

\section{}

\begin{figure}[t]
  \includegraphics[width=3.3in]{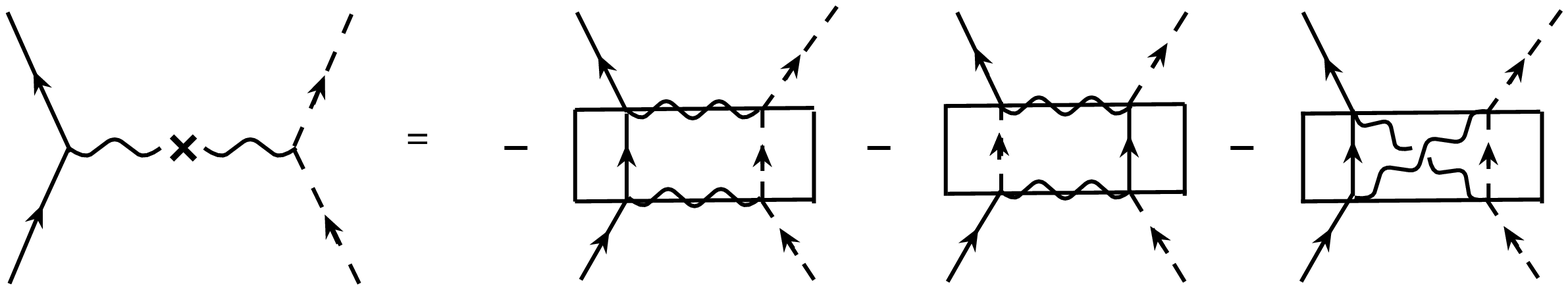}
  \caption{The counterterm diagrams of $\Delta g_{2R}$ up to one-loop order displayed in terms of constituent
  blocks.}\label{counter2}
  \includegraphics[width=2.8in]{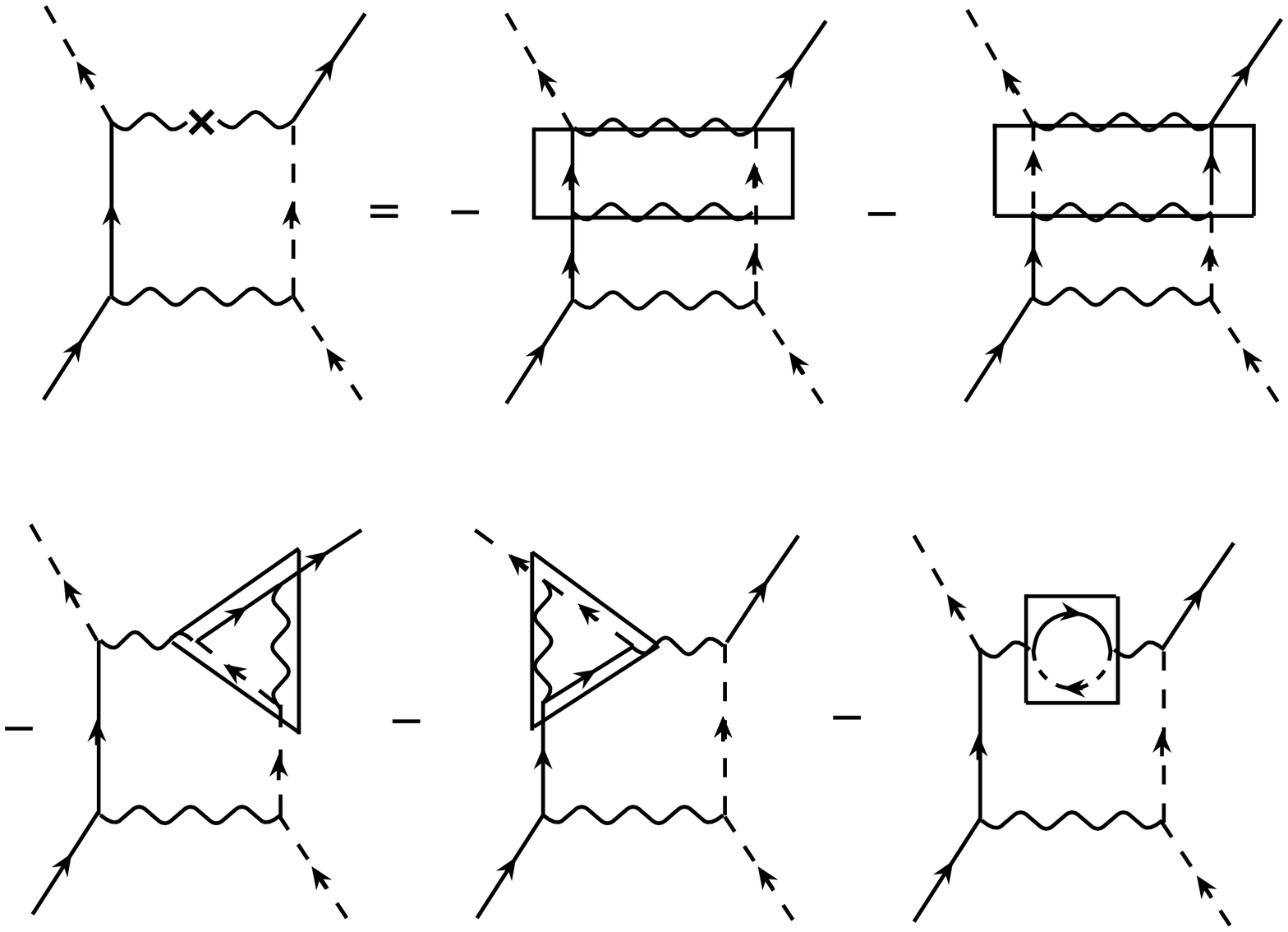}
  \caption{A two-loop example with $\Delta g_{1R}$ mixed with the usual interaction
  vertices.}\label{counter3}
  \includegraphics[width=3.3in]{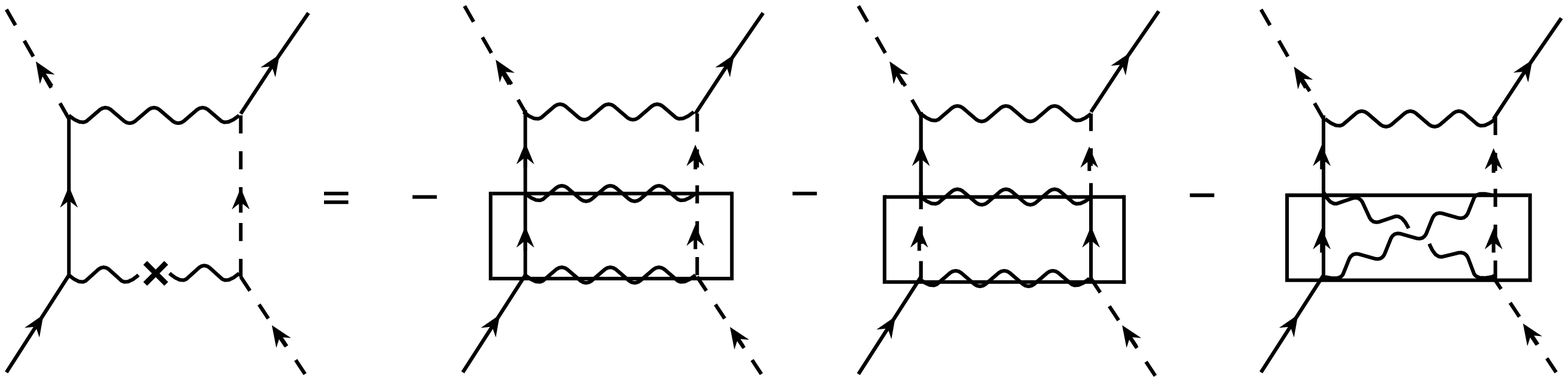}
  \caption{A two-loop example with $\Delta g_{2R}$ mixed with the usual interaction vertices.}\label{counter4}
 \end{figure}

In this appendix, we explain briefly how the cancellation of the
$\ln^2\left(\Omega/\omega\right)$ contributions takes place in
two-loop order within our scheme. In higher-loop orders, only the
cancellations of the leading divergence are guaranteed by the
existing one-loop counterterms. However, we saw that in two-loop
order there were also new diagrams associated with nonleading
logarithmic divergence, which were not accounted by the one-loop
order counterterms. This led us to redefine $\Delta g_{1R}$ and
$\Delta g_{2R}$ in two-loop order. This is consistent with the fact
that we do not know, \emph{a priori}, the exact expression for the
counterterms and, for this reason, they must be calculated order by
order in perturbation theory.

Firstly, recall Eq.(\ref{lag2l}), in which the Lagrangian is written
in terms of the renormalized parameters of the theory together with
the corresponding counterterms. Since every term in the interacting
part of the Lagrangian generates Feynman diagrams in perturbation
theory, so do the counterterms in our case.

\begin{figure}[b]
  \includegraphics[width=1.9in]{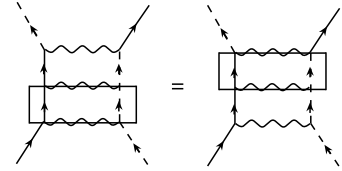}
  \caption{A symmetry relation example of diagrams generated by different constituent blocks.}\label{counter5}
\end{figure}

To illustrate our argument, the counterterms in one-loop order for
$\Delta g_{1R}^{1-loop}$ and $\Delta g_{2R}^{1-loop}$ are now
displayed graphically in terms of constituent blocks in
Figs.\ref{counter} and \ref{counter2}. In this way, when we go to
higher orders, Feynman diagrams are generated, in which these blocks
mix with the usual interaction vertices. This will produce the
expected cancellations of the leading divergence in that order of
perturbation. As an example, consider the diagrams of
Figs.\ref{counter3} and \ref{counter4}. Since the $\Delta
g_{1R}^{1-loop}$ and $\Delta g_{2R}^{1-loop}$ are logarithmically
divergent on their own, when they are inserted in a logarithmically
divergent bubble, they produce a $\ln^{2}\left(\Omega/\omega\right)$
contribution. This cancels exactly the leading order divergence
produced by conventional perturbation theory. However, in order to
avoid double-counting of these counterterms contributions, we need
to take into account the various symmetry relations which exist
among diagrams constructed with some of the counterterms blocks. We
illustrate one of these symmetry relations in Fig.\ref{counter5}.
Following this, only the single logarithmic divergences from the
nonparquet diagrams are left out of these cancellation processes and
they are used to redefine $\Delta g_{1R}$ and $\Delta g_{2R}$ in
two-loop order.

\end{document}